\documentclass[a4paper,11pt]{article}
\pdfoutput=1 

\usepackage{jheppub} 

\usepackage[T1]{fontenc} 

\newcommand \sg{\sqrt{-g}}
\usepackage{changes}
\usepackage{braket}

\title{\boldmath Quantum critical scaling and holographic bound for transport coefficients
	near Lifshitz points}

\author[a]{Gian Andrea Inkof,}
\author[a]{Joachim M.C. K\"uppers,}
\author[a,b]{Julia M. Link,}
\author[c]{Blaise Gout\'eraux,}
\author[a,d]{and J\"org Schmalian}

\affiliation[a]{Institute for Theory of Condensed Matter, Karlsruhe Institute of
	Technology, 76131 Karlsruhe, Germany}
\affiliation[b]{Department
	of Physics, Simon Fraser University, Burnaby, British Columbia,
	Canada V5A 1S6}
\affiliation[c]{CPHT, CNRS, Ecole polytechnique, IP Paris, F-91128 Palaiseau, France}
\affiliation[d]{Institute for Solid State Physics, Karlsruhe Institute of Technology,
	76131 Karlsruhe, Germany}

\emailAdd{gian.inkof@kit.edu}
\emailAdd{julia\_monika\_link@sfu.ca}
\emailAdd{blaise.gouteraux@polytechnique.edu}
\emailAdd{joerg.schmalian@kit.edu}

\abstract{The transport behavior of strongly anisotropic systems is significantly richer compared to isotropic ones. The most dramatic spatial anisotropy at a critical point occurs at a Lifshitz transition, {found in systems with merging Dirac or Weyl point or near the superconductor-insulator quantum phase transition. Previous work found that in these systems a famous conjecture on the existence of a lower bound for the ratio of a shear viscosity to entropy is violated, and proposed a generalization of this bound for anisotropic systems near charge neutrality involving the electric conductivities.}
	The present study uses scaling arguments and the gauge-gravity duality to {confirm the previous analysis of universal bounds in anisotropic Dirac systems.} We investigate the strongly-coupled phase of {quantum Lifshitz} systems in a gravitational Einstein-Maxwell-dilaton model with a linear massless scalar which breaks translations in the boundary dual field theory and sources the anisotropy. The holographic computation demonstrates that some elements of the viscosity tensor can be related to the ratio of the electric conductivities {through a simple geometric ratio of elements of the bulk metric evaluated at the horizon,} and thus obey a generalized bound, while others violate it. From the IR critical geometry, we express the charge diffusion constants in terms of the square butterfly velocities. The proportionality factor turns out to be direction-independent, linear in the inverse temperature, and related to the critical exponents which parametrize the anisotropic scaling of the dual field theory.}

\arxivnumber{1907.05744}

\begin{document} 
\maketitle
\flushbottom

\section{Introduction}
Bounds on transport coefficients are an important tool to quantify
the strength of correlations in quantum many-body systems. If one
can identify a theoretical value for a minimal electrical conductivity
or viscosity, then one can judge
how strongly-interacting a system is. A highly influential bound for
momentum conserving scattering of quantum fluids was proposed by Kovtun,
Son, and Starinets  \cite{Kovtun2005} (KSS) for the ratio of the shear viscosity
and entropy density 
\begin{equation}
\eta/s\geq\frac{\hbar}{4\pi k_{{\rm B}}}.\label{eq:KSS}
\end{equation}
It is obeyed in systems like the quark gluon plasma  \cite{Schaefer2014}
or cold atoms in the unitary scattering limit \cite{Thomas2009}. Graphene
at charge neutrality is another example that is expected to be close
to this bound \cite{Mueller2009}. Within the Boltzmann transport theory
one finds that a bound for $\eta/s$ can be related to the
ratio $l_{{\rm mfp}}/\lambda$ of the mean-free path $l_{{\rm mfp}}$ and the mean distance $\lambda$ between carriers. However, Eq.(\ref{eq:KSS})
is valid even for systems that cannot be described in terms of the
quasi-classical Boltzmann theory. Indeed, the bound is saturated for
quantum field theories in the strong coupling limit as was shown in
Ref.\cite{Kovtun2005} using the holographic duality of conformal
field theory and gravity in anti-de-Sitter spacetime \cite{Maldacena1998,Witten1998,Gubser1998}.
\\
Limiting bounds for the charge transport like the electrical conductivity
are somewhat more subtle. A much discussed example is the Mott-Ioffe-Regel
limit \cite{Ioffe1960,Mott1972,Gurvitch1981} that corresponds to a
threshold value of the electrical resistivity when $l_{{\rm mfp}}/\lambda\sim{\cal O}\left(1\right)$.
While some systems clearly show a saturation of the resistivity once
$\lambda/l_{{\rm mfp}}$ reaches unity, materials like the cuprate
or iron-based superconductors violate this limit \cite{Emery1995}.
For a detailed discussion of correlated materials that obey or systematically
violate the Mott-Ioffe-Regel bound, see Ref.\cite{Hussey2004}. \textcolor{black}{Transport
	properties in quantum critical systems were argued under certain circumstances
	to be governed by a Planckian relaxation rate $\hbar\tau^{-1}\approx k_\text{B}T$ \cite{Zaanen2004,Sachdev2011},
	which would also limit the electrical conductivity
	at quantum critical points.}\textcolor{red}{{} }A bound on charge transport
that is less restrictive and theoretically better justified than the
Mott-Ioffe-Regel limit was proposed in Ref.\cite{Hartnoll2015}.
It constrains the value of the charge diffusivity: 
\begin{equation}
D_{c}\geq C_{D}\frac{\hbar v^{2}}{k_{{\rm B}}T},\label{eq:diffusivity}
\end{equation}
with $C_{D}$ is a numerical coefficient of order unity{. Here} $v$
is a characteristic velocity of the problem. {At charge neutrality the heat and electric currents are decoupled, and the charge diffusivity is determined by the Einstein relation {$D_{c} = \sigma/{\chi}_\rho$.}} $\sigma$ is the electrical
conductivity, and {${\chi}_{\rho}=\partial\rho/\partial\mu$} the charge
susceptibility with particle density $\rho$ and chemical potential
$\mu$. The latter is related to the charge compressibility since
{${\chi}_{\rho}=-\frac{\rho^{2}}{V}\frac{\partial V}{\partial p}$}.  If {$v^{2}{\chi}_{\rho}$} stays constant as $T\rightarrow0$, the electrical
resistivity cannot vanish slower than linearly in $T$ \cite{Hartnoll2015}.
Ref.\cite{Blake2016,Blake2016b} proposed  the butterfly velocity $v=v_{\mathrm{B}}$ as the characteristic velocity.
$v_{\mathrm{B}}$ follows from the analysis of out-of-time-order (OTOC) correlations
$C\left(\mathbf{x},t\right)=-\left\langle \left[A\left(\mathbf{x},t\right),B\left(\mathbf{0},0\right)\right]^{2}\right\rangle $
that are discussed in the context of chaos and information scrambling \cite{Shenker2014,Shenker2015,Roberts2015,Roberts2016,Maldacena2016}.
It can be obtained from the long-distance behavior, e.g. via
\begin{equation}
C\left(\mathbf{x},t\right)\sim e^{2\lambda_\text{L}\left(t-\frac{\mathbf{\left|x\right|}}{v_{\mathrm{B}}}\right)}.
\end{equation}
The scrambling rate $\lambda_\text{L}$ that enters the OTOC is also subject
to the bound $\lambda_\text{L}\leq 2\pi k_\text{B}T/\hbar$ \cite{Maldacena2016}.
While the interpretation of $\lambda_\text{L}$ and its relation to transport
and thermalization rates is not always correct \cite{Blake:2017qgd,Klug2018,Davison:2018ofp,Davison2018,1710.00921,1706.00019},
the butterfly velocity seems to yield a natural scale for the characteristic
velocity of a system, even if no clear quasiparticle description is
available.  A caveat applies when a symmetry of the system is weakly broken and triggers a sound-to-diffusion crossover: in this case, the resulting diffusivity is more naturally expressed in terms of the sound velocity and the gap  \cite{Davison2015b,Davison:2018ofp,Grozdanov:2018fic}.

The focus of this paper is the investigation of anisotropic systems,
where the conductivity tensor $\sigma_{\alpha\beta}$ and the viscosity
tensor $\eta_{\alpha\beta\gamma\delta}$ exhibit a more complex structure
with potentially different temperature dependencies for distinct tensor
elements  \cite{Cook2019,Link2018}. The anisotropy that we consider is most naturally expressed
in terms of the relation between characteristic energies and momenta
along different directions. For a system with two space dimensions, it holds then that:  
\begin{eqnarray}
\omega & \sim & \left|k_{x}\right|^{z/\phi}\nonumber \\
\omega & \sim & \left|k_{y}\right|^{z}\label{eq:def phi 1}
\end{eqnarray}
with dynamical exponent $z$. We characterize the anisotropy
in terms of the exponent $\phi$ that relates typical momenta along
the two directions according to 
\begin{equation}
\left|k_{x}\right|\sim\left|k_{y}\right|^{\phi}.\label{eq:def phi 2}
\end{equation}
{A single particle dispersion that is consistent with such scaling
	would be $\varepsilon\left(\mathbf{k}\right)\sim\left|k_{x}\right|^{z/\phi}+a\left|k_{y}\right|^{z}$
	that corresponds to a system at a Lifshitz point \cite{Lifshitz1942,Dzyaloshinski1964,Goshen1974,Hornreich1975,PhysRevB.62.12338,SHPOT2001340,Shpot_2005,0802.2434}.}
However, our conclusions do not require the existence of well defined
quasiparticles with this dispersion relation. 

{Anisotropic systems, that obey scaling behavior of a Lifshitz transition
	were recently shown to violate the viscosity bound \cite{Rebhan2012,Jain2015, Ge:2014aza,Ge2017,Link2018,1805.01470, Ge:2014aza,1205.1797,1306.1404,Pedraza2018}.}
In Ref.\cite{Link2018} a model of anisotropic Dirac fermions that emerged from two ordinary Dirac cones was analyzed as an explicit condensed matter realization \cite{Isobe2016}. Within a quasiparticle
description of the transport processes and a Boltzmann equation approach,
the conductivity anisotropy was found to diverge: one direction is
metallic and another one insulating. Based on the quasiparticle transport
theory, a modified bound was conjectured, that involves not just the
viscosity tensor elements $\text{\ensuremath{\eta_{\alpha\beta\alpha\beta}}}$
and the entropy density $s\left(T\right)$, but also the conductivities \cite{Link2018}:
\begin{equation}
\frac{\eta_{\alpha\beta\alpha\beta}}{s}\frac{\sigma_{\beta\beta}}{\sigma_{\alpha\alpha}}\geq\frac{\hbar}{4\pi k_{{\rm B}}}.\label{eq:bound}
\end{equation}
Here, no summation over repeated indices is implied.\\
Other tensor elements like $\eta_{\alpha\beta\beta\alpha}$ continue to
obey Eq.(\ref{eq:KSS}). The origin for this combined viscosity-conductivity
bound is the different scaling behavior of the typical velocities
$v_{\alpha}$ for different directions. { Candidate materials with Lifshitz transitions are the organic conductor $\alpha-(\text{BEDT-TTF}_2)\text{I}_3$ under pressure \cite{Julia27}, and the heterostructure of the $5/3 \text{TiO}_2/\text{VO}_2$ supercell \cite{Julia28,Julia29}. Moreover, the surface modes of topological crystalline insulators with unpinned surface Dirac cones \cite{Julia30} and quadratic double Weyl fermions \cite{Julia31} are expected to exhibit such a behavior}. 

The analysis of Ref.\cite{Link2018}
was based on the Boltzmann equation and did not allow to explicitly
analyze a model that satisfies this bound or determine the precise
numerical coefficient in Eq.(\ref{eq:bound}), i.e. the factor ${1}/{4\pi}$.
This can only be done within a formalism that addresses transport
in strongly-coupled non-quasi-particle many-body systems. In the same
context it is of interest to address the related question of whether
the diffusivity bound, Eq.(\ref{eq:diffusivity}), is also modified
for anisotropic systems.

In this paper we perform a holographic analysis of anisotropic transport,
exploiting the duality between strongly coupled quantum field theories
in $d+1$ dimensions and gravity theories in one additional dimension \cite{Maldacena1998}.
{The calculation is based on an Einstein-Maxwell-dilaton (EMD) action, where the anisotropy is generated by massless scalars, linear in  the boundary spatial coordinates. {See Refs. \cite{Rebhan2012,Jain2015,Ge:2014aza,1306.1404,Jeong2018,Donos:2014uba,Ge2017,Mateos:2011ix,Mateos:2011tv,1205.1797,Pedraza2018,Donos2014,1202.4436} for previous studies of these holographic systems.} } {As a consequence, the scalars also break translations {and} 
	momentum is not conserved{. If the symmetry breaking occurs explicitly,} the viscosity cannot be interpreted as a hydrodynamic coefficient.}
{It is well known that in such holographic frameworks the KSS bound is violated \cite{Rebhan2012,Jain2015, Ge:2014aza,Ge2017,1805.01470, Ge:2014aza,1205.1797,1306.1404,Pedraza2018,Hartnoll2016,Alberte:2016xja,Ciobanu:2017fef,Burikham2016,Ling_2016,Ling_2017,1510.06861}.}\\
{We will also investigate the case where translations are broken spontaneously through the use of a so-called Q-lattice homogeneous Ansatz \cite{Amoretti2018_CDW,Amoretti:2017axe,1904.11445}, in which case momentum is still conserved and the shear viscosity remains well-defined at all temperatures.}\\
 {As our focus in this work is on the anisotropy of the system, we
will choose a geometry where momentum is conserved along one of
the spatial directions, say the $\beta$-direction. Thus, the stress tensor
elements $T_{\alpha \beta}$ serves as currents of the conserved momentum density along the direction $\beta$.
Consequently, the viscosity elements $\eta_{\alpha \beta \gamma \beta}$ maintain
their meaning as hydrodynamic coefficients, for all $\alpha$ and $\gamma$.

{We compute the anisotropic electric conductivities in this holographic system, and find that, at charge neutrality,} {their ratio is given by a simple geometric ratio of the spatial elements of the bulk metric evaluated at the bulk black hole horizon. Their temperature dependence indicates metallic behavior along one spatial direction and insulating behavior along the other, as in Ref.\cite{Link2018}.}

{Returning to the viscosities, we find that in the direction where momentum is conserved, the viscosity matrix elements are governed by the same geometric ratio as the electric conductivities, so that, \cite{Link2018}:}
\begin{equation}
\frac{\eta_{\alpha\beta\alpha\beta}}{s} =
\frac{\hbar}{4\pi k_\text{B}} \frac{\sigma_{\alpha\alpha}}{\sigma_{\beta\beta}}.
\label{eq:at the bound anisotr.}
\end{equation}
The generalized bound Eq.\eqref{eq:at the bound anisotr.} has to be understood as a relation between hydrodynamic coefficients{, which holds at all temperatures, including the low temperature regime where the anisotropy is large}. Moreover, the combination $\frac{\eta_{\alpha\beta\alpha\beta}}{s} \frac{\sigma_{\beta\beta}}{\sigma_{\alpha\alpha}}$ serves as an indicator of strong coupling behavior in anisotropic systems.
In Fig.\ref{temperature_behavior} we show typical temperature dependencies for these
transport coefficients for a specific value of the crossover exponent $\phi$ that characterizes the anisotropy. 
{When translations are broken along the $\beta$-direction, $\eta_{\alpha \beta \gamma \beta}$ loses its hydrodynamic meaning and just gives the stress-tensor correlation function. The tensor element satisfies a holographic relation which we analyze in both the limits of high and low temperature.}

In addition, we determine the anisotropic butterfly velocity $v_{B,\alpha}$ ({see Refs. \cite{1805.01470,Jeong2018,Pedraza2018,1610.02669,Blake:2017qgd,1710.05765,1708.07243,1811.06949} for previous studies})
and the compressibility, and obtain for the anisotropic diffusivity
the generalization of Eq.(\ref{eq:diffusivity}) 
\begin{equation}
D_{c,\alpha} =	\frac{d_\text{eff}-\theta}{\Delta_\chi} \frac{\hbar v_{B,\alpha}^{2}}{2\pi k_{{\rm B}}T},
\label{eq:Diffusivity aniso}
\end{equation}
where $d_\text{eff}$ is the effective spatial dimensionality -- see Eq.(\ref{deff}) below, $\theta$ the hyperscaling violating exponent,  and $\Delta_\chi$ the scaling dimension of the charge susceptibility. Thus, the bound of Eq.(\ref{eq:diffusivity}) can be generalized to anisotropic systems. In distinction to the viscosity bound, the anisotropy only changes the universal coefficient that now depends on the exponents $\phi,z,$ and $\theta$. Furthermore, (\ref{eq:Diffusivity aniso}) recovers the limit of isotropic charge neutral theories \cite{Blake2016}. In Ref.\cite{Jeong2018}, the thermal diffusivity was computed in anisotropic setups and also found to obey a relation similar to \eqref{eq:Diffusivity aniso}. See Ref.\cite{1805.01470} for an alternative proposal to \eqref{eq:Diffusivity aniso} at an anisotropic QCP.

Before we present the theories that yield these results, we
give some general scaling arguments, assuming charge and momentum conservation. This analysis motivates us to consider
the appropriate combinations of transport quantities that enter Eq.(\ref{eq:bound})
and Eq.(\ref{eq:Diffusivity aniso}). The scaling analysis is then
followed by a holographic analysis of the combined viscosity-conductivity
bound, the charge susceptibility, and the butterfly velocity within
an anisotropic gravity theory.
\begin{figure}[t]
	\centering
	\includegraphics[scale=0.5]{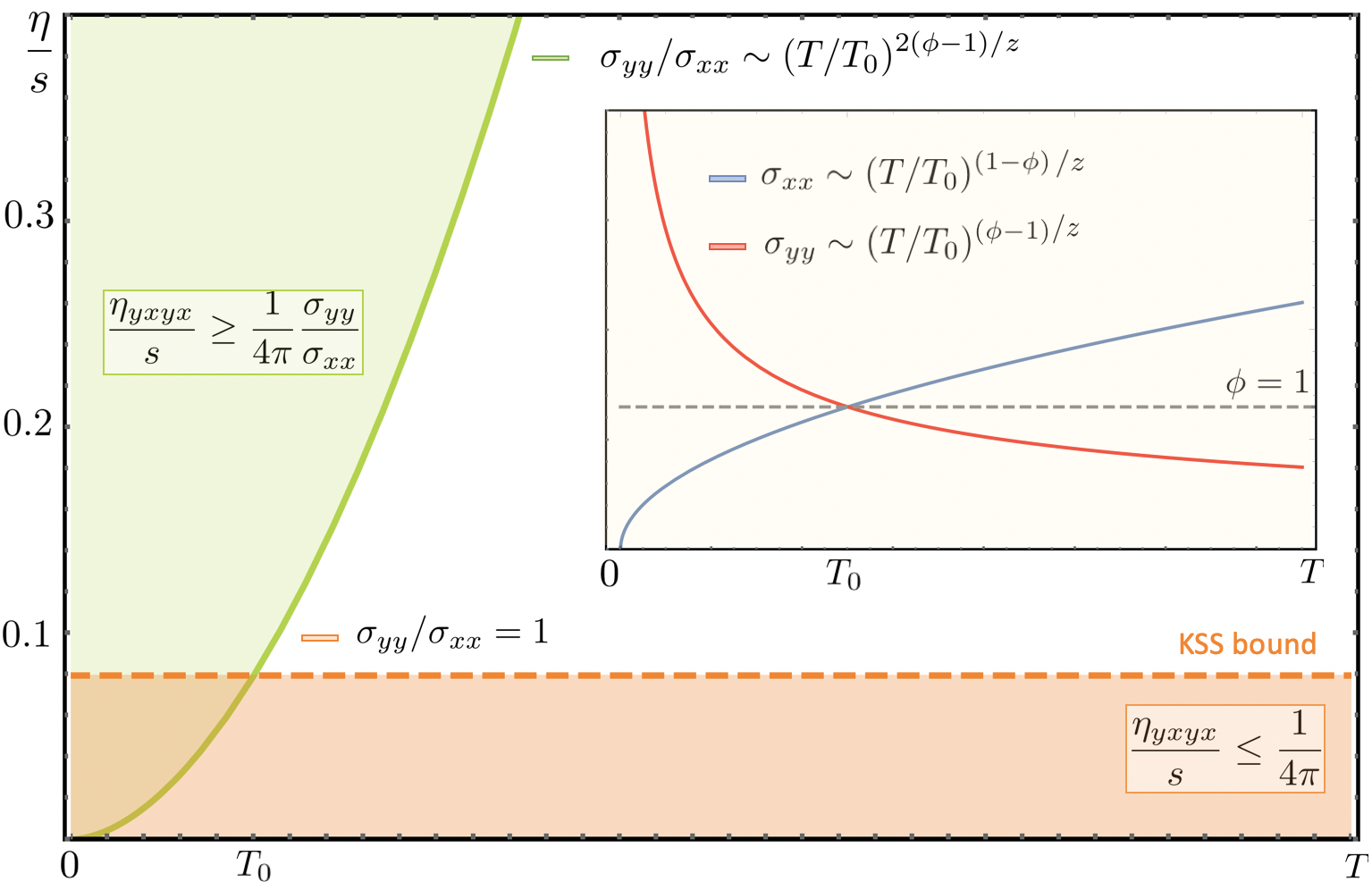}
	\caption{
		Main panel: temperature dependence of the $\eta/s$ tensor. In the anisotropic case the KSS bound (orange) can be parametrically violated (green line). Here, $T_0$ is a temperature scale below which the anisotropy effects are dominant. The conductivity ratio might constitute a new lower bound when rotations are broken (green).
		Inset: temperature dependence of the conductivity tensor elements $\sigma_{xx}$
		and $\sigma_{yy}$.
		$\phi$ is the crossover exponent that characterizes the anisotropy between the different spatial
		directions $k_{x}\sim k_{y}^{1/\phi}$. Once $\phi\protect\neq0$
		one element of the conductivity of a two-dimensional system must be
		insulating and the other must be metallic.
	}
	\label{temperature_behavior}
\end{figure}
\section{Scaling arguments}

\label{scaling_arguments} We consider the scaling behavior of transport
coefficients in anisotropic systems near a quantum critical Lifshitz point.
As we will see, scaling arguments can be efficiently used to make
statements about transport bounds. Once a combination of physical observables has scaling dimension zero, it naturally
approaches a universal value in the limit $T,\mu,\omega\cdots\rightarrow0$,
that corresponds to an underlying quantum critical state. If one can
argue, usually based on an analysis of conservation laws, that
this value is neither zero nor infinity, it should be some dimensionless
number times the natural unit of the observable. In other words, this combination should be insensitive to irrelevant deformations of the quantum critical point. As an example we
consider the electrical conductivity at zero density. For isotropic systems its scaling
dimension is $d-2$, a result that follows from single-parameter scaling
and charge conservation. Thus the conductivity of a zero density two-dimensional
system is expected to reach a universal value in units of the natural
scale $e^{2}/h$. Under the same conditions, both the viscosity and
the entropy density have scale dimension $d$ such that their ratio
has scaling dimension zero. Then $\eta/s$ should approach a universal
value times $\hbar/k_{{\rm B}}$ which yields the correct physical
unit. This observation helps to rationalize a result like Eq.(\ref{eq:KSS}).
As an aside, these scaling considerations also offer a natural explanation
why the bound Eq.(\ref{eq:KSS}), while applicable, is not very relevant
for Fermi liquids. Here, the existence of a large Fermi surface gives
rise to hyper-scaling violating exponents  \cite{Huijse:2011ef}. If
one performs the appropriate scaling near the Fermi surface \cite{Shankar1994},
then it seems more natural to use $\eta s^{2}$ as the natural bound,
a quantity that approaches a constant value as $T\rightarrow0$. 

The conclusions of this section require that scaling relations are
valid, i.e that the system under consideration behaves critical and
is below its upper critical dimension. In the remainder of this section
we assume that this is the case. To be specific, we analyze a $d$-dimensional
system and allow for one direction to be governed by a characteristic
length scale with a different scaling dimension $\phi\neq1$
than the other spatial directions, see Eqs.(\ref{eq:def phi 1},\ref{eq:def phi 2})
above. In addition, the temporal direction is characterized by a dynamic
scaling exponent $z$. Let us then consider a physical observable $O\left(\mathbf{k},\omega\right)$.
By assumption the observable obeys the scaling relation 
\begin{equation}
O\left(k_{\perp},\mathbf{k}_{\parallel},\omega\right)=b^{-\Delta_{O}}O\left(b^{\phi}k_{\perp},b\mathbf{k}_{\parallel},b^{z}\omega\right).\label{eq:scaling law}
\end{equation}
Here $\Delta_{O}$ is the scaling dimension of the observable. The
$d$-dimensional momentum vector $\mathbf{k}=\left(k_{\perp},\mathbf{k}_{\parallel}\right)$
consists of one component $k_{\perp}$ that is governed by the exponent
$\phi$ and a $d-1$ dimensional component $\mathbf{k}_{\parallel}$.
In the subsequent holographic analysis we focus on a system with two
spatial coordinates and use the notation $k_{\perp}=k_{x}$ and $k_{\parallel}=k_{y}$.
While the scaling analysis presented here cannot determine the values
of the exponents, it allows for rather general conclusions once those
exponents are known. For an explicit model with nontrivial exponents
$z$ and $\phi$, see Ref.\cite{Link2018}. 

\subsection{Scaling of thermodynamic quantities}

We begin our discussion of scaling laws with thermodynamic quantities.
For the free-energy density of the system holds the following scaling
law: 
\begin{equation}
F\left(T,\mu\right)=b^{-d_{{\rm eff}}-z}F\left(b^{z}T,b^{z}\mu\right),
\end{equation}
with effective dimension 
\begin{equation}
d_{{\rm eff}}=d-1+\phi.
\label{deff}
\end{equation}
As an energy density, $F$ should scale like unit energy per unit
volume. To obtain its scaling dimension it is then easiest to start
from the usual result $d+z$ for isotropic systems \cite{Sachdev2011}
and replace $d$ by $d_{{\rm eff}}$. This takes into account the
different weight of the directions $\mathbf{k}_{\parallel}$ and $k_{\perp}$.
With $s=-\partial F/\partial T$ and $\rho=\partial F/\partial\mu$
we obtain immediately the scaling dimensions 
\begin{equation}
\Delta_{s}=\Delta_{\rho}=d_{{\rm eff}}
\end{equation}
for the entropy density $s$ and particle density $\rho$, respectively.  Away from zero density, the relation $\Delta_{\rho}=d_{{\rm eff}}$ generally does not hold  \cite{Gouter_2014_mom_diss,Hartnoll2015-1,Davison:2018ofp}.
The second derivative of the free energy with respect to the chemical
potential yields charge susceptibility
{
\begin{equation}
{\chi}_{\rho}\left(T,\mu\right)=b^{-\Delta_{\chi}}{\chi}_{\rho}\left(b^{z}T,b^{z}\mu\right)\label{eq:chi scaling}
\end{equation}
}
with $\Delta_{\chi}=d_{{\rm eff}}-z$. We can now use these thermodynamic
relations to determine the scaling behavior of the conductivity and viscosity. To
do so is possible because of the restrictions that follow from charge and momentum
conservation.

\subsection{Scaling of transport coefficients}

The conductivity is determined via a Kubo formula from the current-current
correlation function, e.g. 
\begin{equation}
{\rm Re}\,\sigma_{\alpha\beta}\left(\omega\right)=\frac{{\rm Im}\,\Pi_{\alpha\beta}\left(\omega\right)}{\omega}.
\end{equation}
At zero density, the system has a finite d.c. conductivity. $\Pi_{\alpha\beta}\left(\omega\right)$
is the Fourier transform of the retarded current-current correlation
function $\Pi_{\alpha\beta}\left(t\right)=-i\theta\left(t\right)\left\langle \left[j_{\alpha}\left(t\right),j_{\beta}\right]\right\rangle $.
In order to exploit the implications of charge conservation we use
the continuity equation 
\begin{equation}
\partial_{t}\rho+\partial_{\alpha}j_{\alpha}=0
\end{equation}
and obtain the well known relation between the longitudinal conductivity
$\sigma_{\alpha\alpha}\left(\omega\right)$ and the density-density
correlation {$\bar{\chi}_{\rho}\left(\mathbf{k},\omega\right)$}
{
\begin{equation}
\sigma_{\alpha\alpha}\left(\omega\right)=\lim_{\mathbf{k}\rightarrow0}\frac{\omega}{k_{\alpha}^{2}}\bar{\chi}_{\rho}\left(\mathbf{k},\omega\right).
\end{equation}
}
Here {$\bar{\chi}_{\rho}\left(\mathbf{k},\omega\right)$} is
the temporal Fourier transform of {$\bar{\chi}_{\rho}\left(\mathbf{k},t\right)=-i\theta\left(t\right)\left\langle \left[\rho\left(\mathbf{k},t\right),\rho\left(-\mathbf{k},0\right)\right]\right\rangle $,}
where $\rho\left(\mathbf{k},t\right)$ is the spatial Fourier transform
of the density $\rho\left(\mathbf{x},t\right)$. Since {${\chi}_{\rho}=\lim_{\mathbf{k}\rightarrow0}\bar{\chi}_{\rho}\left(\mathbf{k},\omega=0\right)$},
the scaling dimension of {${\chi}_{\rho}$} is also $\Delta_{\chi}$,
given below Eq.(\ref{eq:chi scaling}). Thus we find 
\begin{eqnarray}
\Delta_{\sigma,\parallel} & = & \Delta_{\chi}+z-2=d_{{\rm eff}}-2,\nonumber \\
\Delta_{\sigma,\perp} & = & \Delta_{\chi}+z-2\phi=d_{{\rm eff}}-2\phi,
\end{eqnarray}
for the conductivities along the two directions. This yields for the
conductivities:
\begin{eqnarray}
\sigma_{\parallel}\left(T,\omega\right) & = & b^{3-\phi-d}\sigma_{\parallel}\left(b^{z}T,b^{z}\omega\right),\nonumber \\
\sigma_{\perp}\left(T,\omega\right) & = & b^{\phi+1-d}\sigma_{\perp}\left(b^{z}T,b^{z}\omega\right).
\end{eqnarray}
If we return to the isotropic limit, where $\phi=1$, both components
of the conductivity behave the same with usual conductivity scaling
dimension $d-2$. Interestingly, in the anisotropic case, this continues to be the
dimension of the geometric mean $\sqrt{\sigma_{\parallel}\sigma_{\perp}}$.
Distinct scaling exponents for the tensor elements imply a different
temperature dependency of the conductivity for different directions.
Thus, a more insulating behavior along one direction will force the
other direction to be more metallic. For a two-dimensional system,
one direction will have to be insulating and the other then has to
be metallic as long as $\phi\neq1$. Finally, the ratio $\sigma_{\parallel}/\sigma_{\perp}$
of the conductivity is governed by $\Delta_{\sigma,\parallel}-\Delta_{\sigma,\perp}=2\left(\phi-1\right)$,
i.e. 
\begin{equation}
\frac{\sigma_{\parallel}\left(T\right)}{\sigma_{\perp}\left(T\right)}=b^{-2\left(\phi-1\right)}\frac{\sigma_{\parallel}\left(b^{z}T\right)}{\sigma_{\perp}\left(b^{z}T\right)}.\label{eq:sigma ratio}
\end{equation}

We can perform a similar analysis for the viscosity tensor. It is given by a different Kubo formula
\begin{equation}
{\rm Re}\,\eta_{\alpha\beta\gamma\delta}\left(\omega\right)=\frac{{\rm Im}\,\Pi_{\alpha\beta\gamma\delta}\left(\omega\right)}{\omega},
\label{viscosity_scaling_section}
\end{equation}
with $\Pi_{\alpha\beta\gamma\delta}\left(\omega\right)$ the Fourier
transform of the retarded stress-tensor correlation function $\Pi_{\alpha\beta\gamma\delta}\left(t\right)=-i\theta\left(t\right)\left\langle \left[T_{\alpha\beta}\left(t\right),T_{\gamma\delta}\right]\right\rangle $.
Momentum conservation gives rise to the continuity equation for the
momentum density {$g_{\alpha}\equiv T^0_\alpha$}:
\begin{equation}
\partial_{t}g_{\beta}+\partial_{\alpha}T_{\alpha\beta}=0.
\end{equation}
We are considering a system without rotation invariance. In this case
it is important to keep track of the order of the tensor indices as
$T_{\alpha\beta}$ cannot be brought into a symmetric form \cite{Link2018b}. From the continuity equation
for the momentum follows for the viscosity 
\begin{equation}
\eta_{\alpha\beta\gamma\delta}\left(\omega\right)=\lim_{\mathbf{k}\rightarrow0}\frac{\omega}{k_{\alpha}k_{\gamma}}\chi_{\beta\delta}^{\left(g\right)}\left(\mathbf{k},\omega\right),
\end{equation}
with momentum-density correlation function $\chi_{\beta\delta}^{\left(g\right)}\left(\mathbf{k},\omega\right)$,
i.e. the Fourier transform of $\chi_{\beta\delta}^{\left(g\right)}\left(\mathbf{k},t\right)=-i\theta\left(t\right)\left\langle \left[g_{\beta}\left(\mathbf{k},t\right),g_{\delta}\left(-\mathbf{k},0\right)\right]\right\rangle $.
Thus, we only need to know the scaling dimension of $\chi_{\beta\delta}^{\left(g\right)}$
to determine the behavior of the viscosity. The easiest way to obtain
this scaling dimension is to realize that under a boost operation,
a velocity field is thermodynamically conjugate to the momentum density.
A velocity has scaling dimension $z-1$ for the directions along $\mathbf{k}_{\parallel}$
and $z-\phi$ for $k_{\perp}$. To capture all the options we write
this as $z-\varphi_{\alpha}$ where $\varphi_{\alpha}=1$ for all
directions but along $k_{\perp}$ where we have $\varphi_{\alpha}=\phi$.
Thus, it holds 
\begin{equation}
\chi_{\beta\delta}^{\left(g\right)}\left(k_{\perp},\mathbf{k}_{\parallel},\omega\right)=b^{-\Delta_{g,\beta\delta}}\chi_{\beta\delta}^{\left(g\right)}\left(b^{\phi}k_{\perp},b\mathbf{k}_{\parallel},b^{z}\omega\right).
\end{equation}
with $\Delta_{g,\beta\delta}=d_{{\rm eff}}-z+\varphi_{\beta}+\varphi_{\delta}$.
In the Appendix \ref{Alternative_viscosity} we obtain the same behavior from an analysis of strain
generators, following Refs. \cite{Link2018b,Bradlyn2012}. Using $\Delta_{g,\beta\delta}$
allows us to determine the scaling behavior of the viscosity tensor
\begin{equation}
\eta_{\alpha\beta\gamma\delta}\left(T\right)=b^{-\Delta_{\eta,\alpha\beta\gamma\delta}}\eta_{\alpha\beta\gamma\delta}\left(b^{z}T\right)
\end{equation}
with 
\begin{eqnarray}
\Delta_{\eta,\alpha\beta\gamma\delta} & = & \Delta_{g,\beta\delta}+z-\varphi_{\alpha}-\varphi_{\gamma}\nonumber \\
& = & d_{{\rm eff}}-\varphi_{\alpha}+\varphi_{\beta}-\varphi_{\gamma}+\varphi_{\delta}.\label{eq: viscosity dimension}
\end{eqnarray}
For isotropic systems, this gives the well known result that the scaling
dimension of the viscosity is $d$, i.e. the same as for the entropy
or particle density. For an anisotropic system the scaling dimensions
of the viscosity and the entropy density can still be the same. This
is the case whenever $\varphi_{\alpha}+\varphi_{\gamma}=\varphi_{\beta}+\varphi_{\delta}$.
Examples are $\eta_{\perp\perp\perp\perp}$, $\eta_{\perp\perp cd}$,
$\eta_{ab\perp\perp}$$,$$\eta_{a\perp\perp d}$, or $\eta_{\perp bc\perp}$,
where $a$,$b$ etc. stand for components of $\mathbf{k}_{\parallel}.$

However, the scaling dimension of the viscosity can also be different
from the one of the entropy density. This is the case for 
\begin{eqnarray}
\eta_{a\perp c\perp}\left(T\right) & = & b^{-\left(d-3+3\phi\right)}\eta_{a\perp c\perp}\left(b^{z}T\right)\nonumber \\
\eta_{\perp b\perp d}\left(T\right) & = & b^{-\left(d+1-\phi\right)}\eta_{\perp b\perp d}\left(b^{z}T\right).
\end{eqnarray}
If we now take the ratio of the viscosity to entropy density, we find
\begin{eqnarray}
\frac{\eta_{a\perp c\perp}\left(T\right)}{s\left(T\right)} & = & b^{-2\left(\phi-1\right)}\frac{\eta_{a\perp c\perp}\left(b^{z}T\right)}{s\left(b^{z}T\right)},\nonumber \\
\frac{\eta_{\perp b\perp d}\left(T\right)}{s\left(T\right)} & = & b^{2\left(\phi-1\right)}\frac{\eta_{\perp b\perp d}\left(b^{z}T\right)}{s\left(b^{z}T\right)}.\label{eq:eta ratio}
\end{eqnarray}
Thus, for $\phi\neq1$ there is always one tensor element of the viscosity,
where $\eta_{\alpha\beta\gamma\delta}/s$ diverges as $T\rightarrow0$
and another one that vanishes. The latter will then obviously violate
any bound for the ratio of a viscosity to entropy density. In Ref.\cite{Link2018}
it was shown that precisely these tensor elements turn out to be important
for the hydrodynamic Poiseuille flow of anisotropic fluids.

The origin of unconventional scaling of both the conductivities and the
viscosities is geometric, i.e. rooted in the anisotropic scaling of
spatial coordinates at the Lifshitz point. If one combines Eqs.(\ref{eq:sigma ratio})
and (\ref{eq:eta ratio}), it is straightforward to see that the combinations
that enter Eq.(\ref{eq:bound}) always have scaling dimension zero.
While it certainly does not offer a proof of Eq.(\ref{eq:bound})
this is necessary for such quantity to approach a universal, constant
low-temperature value.

Finally we comment on the scaling behavior of the diffusivity bound,
Eq.(\ref{eq:Diffusivity aniso}). To check whether this bound even
makes sense for an anisotropic system, we consider the quantity 
\begin{equation}
X_{\alpha}=k_{{\rm B}}TD_{c,\alpha}/\hbar v_{\alpha}^{2}
\end{equation}
where $v_{\alpha}$ is the characteristic velocity along the $\alpha$-th
direction and {$D_{c,\alpha}=\sigma_{\alpha\alpha}/{\chi}_{\rho}$} the diffusivity
along this direction. It obviously holds 
\begin{equation}
\Delta_{X_{\alpha}}=z+\Delta_{\sigma,\alpha}-\Delta_{\chi}-2\left(z-\varphi_{\alpha}\right)
\end{equation}
where we used again that a velocity scales as $z-\varphi_{\alpha}$.
If we now insert our above results, it follows 
\begin{equation}
\Delta_{X_{\parallel}}=\Delta_{X_{\perp}}=0.
\end{equation}
This implies that $X_{\alpha}$ should approach a universal constant
times $\hbar/k_{{\rm B}}$. Thus, we expect Eq.(\ref{eq:diffusivity})
to be valid even for anisotropic systems, which yields Eq.(\ref{eq:Diffusivity aniso}).
In this sense is this bound even more general than the original viscosity
bound of Eq.(\ref{eq:KSS}).

\section{Holographic analysis of the viscosity-conductivity bound}

The correspondence between gravity theories and quantum field theories,
as it occurs in the anti-de Sitter space/conformal field theory
duality \cite{Maldacena1998,Witten1998,Gubser1998}, is a powerful
tool to analyze the universal properties of strongly-coupled field theories.
In what follows we analyze an anisotropic bulk geometry in order to
determine the relationships between distinct transport coefficients
of anisotropic quantum many-body problems in the strong-coupling limit.
To this end we use the membrane paradigm \cite{Thorne1986} to express
boundary theory transport coefficients in terms of geometric quantities
at the horizon \cite{Iqbal2009}. To be specific, we consider a system
of two space dimensions, i.e. with $D=2+1$ space-time coordinates
at the boundary. The relation between the generating functional of
the quantum field theory and the gravity action for imaginary time is given by \cite{Witten1998,Gubser:1998bc}
\begin{equation}
\left\langle e^{-\int d^{3}x\Phi_{0}{ O}}\right\rangle =\left.e^{-S\left[\Phi\right]}\right|_{\Phi\left(r\rightarrow0\right)=\Phi_{0}},
\end{equation}
where ${ O}$ is an operator of the field theory, $\Phi_{0}$
a conjugate source, $\Phi$ the dual field, and $S$ a gravitational action in
the $D+1$ dimensional bulk, with additional coordinate $r$. Here
we chose a system of coordinates where the boundary lies at
$r=0$. Following Ref.\cite{Iqbal2009}, retarded Green's functions
of the field theory can be obtained from 
\begin{equation}
\left\langle { O}\left(\mathbf{x},t\right)\right\rangle _{\Phi_{0}}=\lim_{r\rightarrow0}\Pi\left(r,\mathbf{x},t\right),
\label{radial_momentum}
\end{equation}
where $\Pi$ is the canonical momentum conjugate to $\Phi$, as follows
from the gravitational version of the Hamilton-Jacobi formalism. This
finally allows for the determination of retarded Green's functions 
$
G\left(\mathbf{x},t\right)=-i\theta\left(t\right)\left\langle \left[{ O}\left(\mathbf{x},t\right),{ O}\left(\mathbf{0},0\right)\right]\right\rangle .
$
For the Fourier transform with respect to momentum and frequency follows
\begin{equation}
G\left(\mathbf{k},\omega\right)=-\lim_{r\rightarrow0}\frac{\Pi\left(r,\mathbf{k},\omega\right)}{\Phi\left(r,\mathbf{k},\omega\right)}.
\end{equation}
Causality is preserved if one considers $\Phi$ that satisfies in-falling
boundary conditions at the black hole horizon \cite{Son2002,Herzog2003}.
The related transport coefficient is given by $-\lim_{\omega,\mathbf{k}\to0}\frac{1}{\omega}{\rm ImG\left(\mathbf{k},\omega\right)}.$

Anisotropic, static bulk geometries cannot come from a pure gravitational
action. Thus, we need to couple gravity with {axial} {gauge} fields or
massless scalar fields. See also Ref.\cite{1202.4436,1310.6725,1306.1404} for generalities on anisotropic studies in holography.
\\
We start from the Einstein-Maxwell-dilaton
action
\begin{equation}
S=\int d^{3+1}x\sqrt{-g}\left(R+{\cal L}_{\text{M}}\right),
\label{EMD_model}
\end{equation}
with Lagrangian 
\begin{equation}
{\cal L}_{\text{M}}= 
-
\frac{1}{2}\left(\nabla\varphi\right)^{2}-V\left(\varphi\right)
-\frac{Y\left(\varphi\right)}{2}\left(\nabla\psi\right)^{2}
-\frac{Z\left(\varphi\right)}{4}F^{2}.
\label{matter_lagrangian}
\end{equation}
$\varphi$ is referred to as the dilaton. It is a scalar field which enters the action modifying
all the couplings involved. $V\left(\varphi\right)$ is its own potential.
We include the dilaton as it will {allow} us to consider anisotropic geometries that arise near the horizon of near-extremal black holes. In the absence of the dilaton
{field, the model} reduces to the usual $\text{AdS}_4$ system with electromagnetic
field, i.e $V\left(0\right)=2\Lambda$ with cosmological constant
$\Lambda=-3/\ell^{2}$, $Z\left(0\right)=1$ and $Y(0)=0$. $\ell$ is the radius of
curvature of the AdS space.
As we will shortly see, by considering a bulk profile that depends linearly on one of the boundary spatial coordinates, the massless scalar $\psi$ will break the rotation and translation symmetries of the dual field theory. {For related work on this family of holographic models, see Refs. \cite{Mateos:2011ix,Mateos:2011tv,Donos:2013eha,Andrade2014,Rebhan2012, Pedraza2018,Ge2017,Ge:2014aza,Jain2015,Ling_2016,Ling_2017,Amoretti2016_chas,Amoretti2018_CDW,Andrade2014,Davison2015b,Donos2014,Hartnoll2016} }
\\
$F^{2}$ is
the Maxwell Lagrangian with $F_{\mu\nu}=\nabla_{\mu}A_{\nu}-\nabla_{\nu}A_{\mu}$
the usual field tensor with vector potential $A_{\mu}$. This term
is needed to implement a $U\left(1\right)$ global symmetry in the
boundary theory and to determine the electrical conductivity.

We summarize the field equations of motion that follow from Eq.\eqref{EMD_model} varying the action with respect to the fields $g_{\mu\nu},A_\mu,\psi,\varphi$.
\\
Varying the metric, we obtain the Einstein equations
\begin{equation}
\label{eq:Einstein}
R_{\mu\nu}-\frac{1}{2}Rg_{\mu\nu}=-\frac{1}{\sqrt{-g}} \frac{ \delta (\sqrt{-g} \mathcal{L}_\text{M})}{\delta g^{\mu\nu}}.
\end{equation}
The variation of the gauge field yields the Maxwell equations
\begin{equation}
\partial_{\mu}(\sqrt{-g}Z(\varphi)F^{\mu\nu})=0.
\label{gaugeDG}
\end{equation}
Notice that both the scalar fields are neutral such that the Maxwell equations are bulk conservation equations for the two-form $F$. Ultimately, this will let us evaluate the boundary charge current at the black hole horizon. Finally, the wave equations for the two scalars are:
\begin{eqnarray}
&\partial_\mu (\sg\,Y(\varphi)\partial^\mu\psi)=0,
\label{axions_EOM}
\\
&\partial_\mu(\sg\,\partial^\mu \varphi)={\partial_\varphi} V_\text{eff},
\label{dialton_EOM}
\end{eqnarray}
where 
\begin{equation}
\frac{V_\text{eff} }{\sg}=
{V(\varphi)+\frac{Y(\varphi)}{2}(\partial\psi)^2+\frac{Z(\varphi)}{4}F^2}.
\label{Veff}
\end{equation}
In the absence of external perturbations, we use the following ansatz
\begin{eqnarray}
&& ds^2 = - g_{tt}(r)dt^2 + g_{rr}(r)dr^2+ \sum_\alpha g_{\alpha\alpha}(r)dx_\alpha^2  \nonumber\\
&&\varphi=\varphi(r), \quad A=A_t(r) dt, \quad \psi = a y,
\label{eq:metric}
\end{eqnarray}
where $a$ is real and $\alpha=\{x,y\}$. The Ansatz for $\psi$ is consistent with the field equations and
preserves the homogeneity of the other fields. Indeed, $\psi$ back-reacts on the equations of motion only through gradients so that all dependence on $y$ drops out of the field equations. However, translations along the $y$-direction are broken and momentum is dissipated at a strength set by $a$. On the other hand, momentum along $x$ direction is conserved which allows us to perform a hydrodynamic analysis of the viscosity tensor elements $\eta_{\alpha x\beta x}$. 
The metric in (\ref{eq:metric}) describes anisotropic bulk geometries since, in general, $g_{xx}\left(r\right)\neq g_{yy}\left(r\right)$. The coefficient $a$ determines the temperature scale $T_0$ below which the anisotropy effects are large. Setting $a=0$ restores both rotations and translations in the dual field theory.
\begin{figure}[t]
	\centering \includegraphics[width=3.4in]{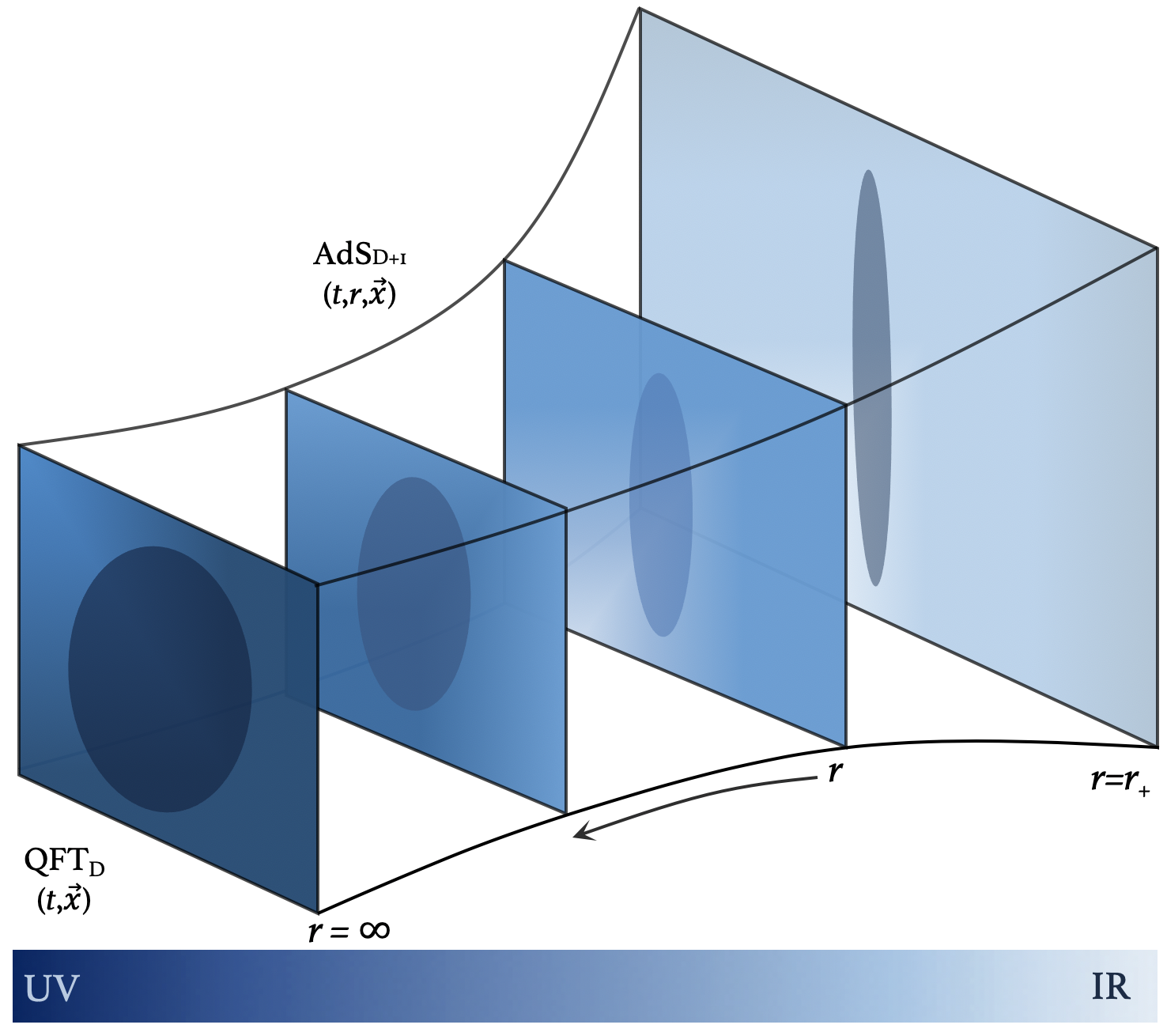} \caption{Cartoon of an AdS black hole geometry in ($D$+1)-spacetime dimensions.
		The extra coordinate $r$ is between $r=0$ and $r=r_{+}$ where the
		boundary and the horizon of the AdS are located, respectively. The
		evolution along $r$ is dual to the RG-flow of the dual $D$-spacetime
		dimensional QFT \cite{McGreevy}. Furthermore, in
		order to underline how the spatial anisotropy becomes higher as one
		approaches the IR region, we have depicted how an ellipse is distorted
		in the spatial directions as $r$ decreases.
	}
	\label{fig:example} 
\end{figure}

In its more general formulation, the holographic correspondence maps the RG-flow of the dual (strongly coupled) field theory to the evolution along the radial direction \cite{McGreevy} -- see Fig.\ref{fig:example}.
The near boundary region
captures the UV of the dual field theory, while the near
horizon region describes the IR. 
In the UV ($r\to0$) the geometry is assumed to be asymptotically $\text{AdS}_4$ : 
	\begin{equation}
		ds^2=  \frac{1}{r^2}\left(  -dt^2 + \frac{dr^2}{r^2} + dx^2+dy^2\right) +\dots
	\label{AdS_metric}
	\end{equation}
where the dots denote subleading terms as $r\to 0$.
This requires
\begin{equation}
\begin{split}
V_\text{UV}\equiv V(0) = -6, 
&\quad
Y_\text{UV}\equiv Y(0)= 0,		\\
Z_\text{UV}\equiv &Z(0) = 1
\end{split}
\end{equation}
with the dilaton vanishing like $\varphi= \varphi_{s} r^{3-\Delta_\varphi}+\varphi_{v} r^{\Delta_\varphi}+\dots$ coming from the near boundary expansion of Eq.(\ref{dialton_EOM}). {$\Delta_\varphi<3$ is the largest solution of $M^2=\Delta_\varphi(\Delta_\varphi-3)$,  $M$ being the mass of the field. The dilaton field is thus dual to a relevant deformation of the UV CFT, with source $\varphi_s$ and vacuum expectation value $\varphi_v$ \cite{McGreevy}. 
	
	From this discussion, we also see that the bulk field $\psi$ sources a marginal deformation of the UV CFT.
	Similarly, $A_{t}=\mu-\rho r+\dots$  with chemical potential $\mu$ and charge density $\rho$. In the following we analyze the charge neutral case $A_t=0$.
	
	Since both scalars are dual to relevant/marginal deformations of the UV CFT, we expect the system to be able to flow to a non-trivial quantum critical phase {in the IR}. This IR endpoint of the RG flow is represented in the bulk by a power law geometry, which arises in the near horizon region at very low temperatures compared to the sources of the UV CFT. To find such geometries, we assume that the dilaton runs logarithmically in the IR {($r\to+\infty$)} $\varphi=2\kappa \log(\hat r/L)$, {where $\hat r$ is an appropriate IR radial coordinate. It is valid in the regime $\hat r\gg L$, where $L$ is the length scale at which the spacetime is well-approximated by its IR scaling form, see \eqref{IR_scaling_model} below. It is generally distinct from the coordinate $r$, which covers all of spacetime}. In the region $\hat r\gg L$, the scalar potentials take the following form \cite{Charmousis:2010zz,Gouter_2014_mom_diss}
	\begin{equation}
	V_\text{IR} = -V_0 e^{\delta \, \varphi}, 
	\quad
	Y_\text{IR} = e^{\lambda \varphi},		
	\quad
	Z_\text{IR} = \, e^{\zeta \varphi}\,.
	\label{coupling_IR}
	\end{equation}
	The critical scaling of the previous section is holographically realized by a {hyperscaling-violating} Lifshitz geometry of the form
	\begin{equation}
	ds^2 = \hat r^\theta \left( -\frac{dt^2}{\hat r^{2z}} +L^2 \frac{d\hat r^2}{\hat r^2} + \frac{dx^2}{\hat r^{2\phi}} + \frac{dy^2}{\hat r^2} \right),
	\label{IR_scaling_model}
	\end{equation}
	which is covariant under the scale transformation
	$
	(t,\hat r,x,y) \to (b^{-z} t,b^{-1}{\hat r}, b^{-\phi} x,b^{-1} y),
	$
	up to a conformal factor $ds^2\to b^{-\theta}ds^2$. Therefore, $\phi$ and $z$ coincide with the anisotropic and dynamical exponents, and $\theta$ quantifies the violation of scale invariance in the metric  \cite{Pedraza2018,Gouteraux:2011ce,Huijse:2011ef}. 
	All the parameters involved are real and $V_0,\delta,L>0$. 
	The explicit derivation of such a solution can be found in Appendix \ref{single_axion_model}, for the (marginally) relevant single massless scalar case, which has $z=\phi\neq1$, the marginally double massless scalars case, which has $z>1$, $\phi\neq1$, and the irrelevant single massless scalar case, which has $z=1$, $\phi=1$ (and where rotations/translations along $x$ are only broken away from the IR endpoint through the irrelevant deformation).
	\\
	A finite temperature can be introduced via the emblackening factor\footnote{We observe that this is an exact solution only when the massless scalar sources only a marginal deformation in the IR. Otherwise the interplay between the irrelevant deformation and temperature is more complicated, although the scaling relation between the location of the event horizon and the temperature still holds.}
	\begin{equation}
	ds^2 = \hat r^\theta \left( -f\frac{dt^2}{\hat r^{2z}} +L^2 \frac{d\hat r^2}{f\hat r^2} + \frac{dx^2}{\hat r^{2\phi}} + \frac{dy^2}{\hat r^2} \right),\quad f(\hat r)= 1-\left({\frac{\hat r}{\hat r_+}}\right)^{\delta_0},
	\label{blackening_factor}
	\end{equation}
	where $\hat r_+$ denotes the location of the event horizon and $\delta_0=1+\phi+z-\theta$.
	The Hawking temperature is 
	\begin{equation}
	4\pi T ={\frac{|\delta_0|}{L}} \hat r_+^{-z}
	\end{equation}
	and satisfies $T\to b^z T$, consistently with the scaling analysis.	
	The fact that scaling stops at a finite value of the flow is reflected
	in the event horizon at finite $\hat r_+$. 
	The entropy density follows from the area of the horizon $s =4\pi \hat r_+^{\phi+1}$. 
	
	Thus, with an appropriate choice of $V\left(\varphi\right)$,
	and $Y\left(\varphi\right)$ we can ``engineer'' a holographic
	dual that generates a desired crossover exponent $\phi$. Without
	more constructive statements about the field theory-gravity dual,
	it is not possible to determine the values of $\phi$ for a given
	quantum field theory. However, we can make statements about a number
	of physical observables for a given value of $\phi$.
	
	\subsection{Analysis of the conductivity}
	In this section we review the results of Ref.\cite{Iqbal2009,Donos2014,Donos:2014uba} to express the electric conductivities in terms of IR quantities. In particular, we calculate the d.c. conductivity along the $\alpha$-direction 
	\begin{equation}
	\sigma_{\alpha\alpha} =
	\lim_{\substack{\omega\to0\\r\to0}} \text{Im}\frac{j^\alpha(r,\omega)}{\omega A_\alpha(r,\omega)},
	\label{electric_conductivity}
	\end{equation}	
	working directly at zero frequency and switching on a constant and small electric field. $A_\alpha$ is the fluctuation respect to which we linearize the gauge equations, and $j^\alpha$ is the associated canonical momentum.
	
	Within the homogeneous ansatz (\ref{eq:metric}), Maxwell equations assume the form
	\begin{equation}
	\partial_{r}(\sqrt{-g}Z(\varphi)F^{\mu r})=0.
	\label{maxwell_conserving}
	\end{equation}
	The quantity in brackets coincides with the conjugate
	momentum of the gauge field $j^{\mu}=\delta S/\delta\left(\partial_{r}A_{\mu}\right)$.
	From the holographic dictionary (\ref{radial_momentum}) follows that the boundary value $j^{\mu}\left(r=0 \right)$ of this quantity is the electric current density of the dual field theory. From (\ref{maxwell_conserving}) $j^\mu$ is radially conserved, i.e. $\partial_{r}j^{\mu}=0$. Thus, we can
	determine the current at the boundary from the behavior of $j^{\mu}$
	at the horizon
	\begin{equation}
	j^{\mu}\left(r=0,\mathbf{x},t\right)=\lim_{r\rightarrow r_{+}}j^{\mu}\left(r,\mathbf{x},t\right).
	\end{equation}
	In the absence of external fields, the only non zero component of
	$j^{\mu}$ is the temporal one $	j^{t}=\sqrt{-g}Z\left(\varphi\right)F^{tr
		\label{charge_density_gauge_field}}
	$
	which corresponds to the charge density $\rho$ of the field theory. In the following we focus on the charge neutral case $\rho=0$.
	\\
	In order to determine the conductivity, we add a small electric field
	$E_{\alpha}=F_{\alpha t}$ in the $\alpha$-direction, e.g. via
	\begin{equation}
	A_{\alpha}^{{\rm ext}}=-E_{\alpha}t.
	\end{equation}
	This electric field will polarize the system and therefore induce
	small corrections to the metric and matter fields. We parametrize
	those corrections via 
	\begin{eqnarray}
	&&A_{\alpha}  =  -E_{\alpha}t+\delta A_{\alpha}\left(r\right),\nonumber \\
	&&	g_{t\alpha}  =  \delta g_{t\alpha}\left(r\right),\nonumber \\
	&&	g_{r\alpha}  =  g_{\alpha\alpha}\left(r\right)\delta h_{r\alpha}\left(r\right),
	\end{eqnarray}
	and $	\psi  =  a y+\delta\psi \left(r\right)$ if $\alpha=y$.
	
	All terms $\delta A_{\alpha}$ etc. are assumed to be of first order
	in the electric field. They can be related to each other through a
	perturbative solution of the field equations.
	The above ansatz yields at first order and for zero density $\rho=0$:
	\begin{equation}
	j^{\alpha}=
	-\frac{\sqrt{-g}Z\left(\varphi\right)}{g_{rr}g_{\alpha\alpha}}\partial_{r}\delta A_{\alpha}.
	\end{equation}
	This result further simplifies our analysis as we only need to determine
	$\delta A_{\alpha}$. To this end we perform a transformation to a
	set of coordinates that is free of singularities at the horizon. This
	is accomplished by the Eddington-Finkelstein (EF)
	coordinates \cite{Eddington1924,Finkelstein1958} $t'=t+r_\star(r)$, where $dr_\star=dr/\gamma(r)$ is the tortoise coordinate, $\gamma(r)=\sqrt{g_{tt}(r)/g_{rr}(r)}$. 
	In these variables holds that near the horizon 
	\begin{equation}
	A_{\alpha}=-E_{\alpha}t'+{E_{\alpha}} r_\star(r)+\delta A_{\alpha}.
	\end{equation}
	If we now demand regularity of $ A_{\alpha}$ in the EF coordinates
	it follows for the leading, singular contribution: 
	\begin{equation}
	\delta A_{\alpha}\left(r\rightarrow r_+\right)=-E_{\alpha} r_\star(r).
	\end{equation}
	It is now straightforward to determine the conductivities 
	\begin{equation}
	\sigma_{\alpha\alpha}=\lim_{r\rightarrow r_+}\frac{j^{\alpha}}{E_{\alpha}}=\sqrt{\frac{g_{\overline{\alpha}\overline{\alpha}}}{g_{\alpha\alpha}}}Z\left(\varphi\right)\bigg|_{r_+},\label{eq:conduct holo}
	\end{equation}
	where $\overline{x}=y$ and $\overline{y}=x$. 
	
	\subsection{Analysis of the viscosity}
	\label{analysis_viscosity}
	In order to compute the correlation function (\ref{viscosity_scaling_section}), we act on the bulk-metric field which is dual to the boundary stress tensor \cite{Kovtun2005}. To get the shear viscosity components, we  switch on small off-diagonal fluctuations of the spatial sector
	\begin{equation}
	ds^2 \mapsto ds^2 + e^{-i \omega t} \delta h_{xy}(r) dxdy.  
	\end{equation}
	In the following we linearize Einstein equations with respect to the one-index-up parametrization $h_\alpha^\beta = g^{\beta	\beta} \delta h_{\alpha\beta}$ and compute the viscosity through
	\begin{equation}
	\eta_{\alpha\beta\alpha\beta}= \lim_{\substack{\omega\to0\\r\to0}} \frac{1}{\omega}\, \text{Im}\frac{\Pi_{\beta}^\alpha(r,\omega)}{ {h}_\alpha^\beta(r,\omega)},
	\label{viscosity_first_step_nodiss}
	\end{equation}	
	where $\Pi_\beta^\alpha$ is the associated radial momentum\footnote{Strictly speaking, the most natural notation in Eq.\eqref{electric_conductivity} and \eqref{viscosity_first_step_nodiss} would have been $\sigma^{\alpha\alpha}$ and $\eta^{\alpha\,\alpha}_{\,\,\beta\,\,\beta}$. Since however the boundary is flat, we could choose the notation with all indices down in order to be consistent with section \ref{scaling_arguments}.}. Since the model is anisotropic, there will be two fluctuations satisfying different equations of motion. 
	
	We start with the simpler case to review the standard derivation of the viscosity, and consider $\delta h_{xy} = g_{xx} h^{x}_{y}$. The Einstein equations (\ref{eq:Einstein}) yield
	\begin{equation}
	\partial_\mu\left( \frac{\sg}{\mathcal{N}} \partial^\mu h^{x}_y \right) =0,
	\end{equation}
	which describes the dynamics of a massless scalar with radial dependent coupling $\mathcal{N}(r) = g_{yy}(r) g^{xx}(r)$. 
	The canonical momentum is 
	\begin{equation}
	\Pi^y_x  = \frac{\sg}{\mathcal{N}}\partial^r  h^x_y,
	\end{equation}
	satisfying ${\mathcal{N}} \partial_r \Pi^y_x = - \omega^2 {\sg} h^x_y$.
	In the low frequency limit, i.e. $\omega\to0$ keeping ${\omega} h^x_y$ and $\Pi$ fixed \cite{Iqbal2009}, both the fluctuation and the momentum are radially conserved allowing to perform a near horizon limit in Eq.(\ref{viscosity_first_step_nodiss}). Here the fluctuation satisfies the in-falling conditions
	\begin{equation}
	h^x_y(r,\omega) \to {h}_0(r)\, e^{-i\omega r_\star(r)}.
	\label{horizon_condition_fluct_nodiss}
	\end{equation}
	$h_0$ is the real solution to the frequency independent wave equation, which asymptotes to a constant at the boundary and is regular at the horizon. Due to the radial conservation $h_0(r)=\text{const}\equiv1$. 
	We then obtain  
	\begin{equation}
	\frac{\eta_{yxyx}}{s} =  \frac{1}{4\pi} \frac{g_{xx}}{g_{yy}}\bigg|_{r_+},  
	\label{viscosity_conserving}
	\end{equation}	 
	which reproduces the bound of Eq.(\ref{eq:KSS}) in the isotropic
	limit $g_{xx}=g_{yy}$. These results, together
	with our findings of Eq.(\ref{eq:conduct holo}) for the conductivities
	immediately yield the expression Eq.(\ref{eq:at the bound anisotr.})
	given in the introduction. 
	
	For the $y$-index-up parametrization we find
	\begin{equation}
	\partial_\mu	\left({{\sg}{\mathcal{N}}\,  \partial^\mu h^{y}_x} \right)
	=
	{\sg}{\mathcal{N}}\, m^2h^{y}_x
	\label{EOMdiss}
	\end{equation}
	with {radially-dependent mass of the shear graviton} $m^2(r)=  a^2Y(\varphi) g^{yy}(r)$ arising due to the breaking of translations along $y$. As before, we define the conjugate momentum via
	\begin{equation}
	\Pi^x_y  = {{\sg}{\mathcal{N}}\,  \partial^r h^{y}_x}, 
	\label{hamilton_eqs_fluctuations_diss}
	\end{equation}
	with $\partial_r \Pi^x_y ={\sg} \mathcal{N}  ( m^2-\omega^2 g^{tt}) h^{y}_x$.
	The non vanishing mass makes the evolution along $r$ non trivial even at zero frequency.
	However, from the equations follows that $\text{Im}\left[\Pi^x_y \,{h}^{y\star}_x\right]$ is radially conserved \cite{Jain2015}, ${h}^\star$ denoting the complex conjugate fluctuation. In particular we can switch to the near horizon limit in the numerator
	\begin{equation}
	\eta_{xyxy} = \lim_{\omega\to0} \frac{ \lim_{r\to r_+} \text{Im}\left[\Pi^x_y \,{h}^{y\star }_x\right]}{ \lim_{r\to 0} \omega|h^{y}_x|^2}.
	\end{equation}	
	Using the in-falling conditions in the numerator we obtain
	\begin{equation}
	\frac{\eta_{xyxy}}{s} =  \frac{1}{4\pi} \frac{g_{yy}}{g_{xx}}\bigg|_{r_+}
	h_0^2(r_+)
	\label{viscosity_diss}
	\end{equation}
	so that the viscosity-conductivity ratio that appears in the conjectured bound \eqref{eq:bound} becomes
	\begin{equation}
	\frac{\eta_{xyxy}}{s}\frac{\sigma_{yy}}{\sigma_{xx}} =  \frac{1}{4\pi}\bigg|_{r_+}
	h_0^2(r_+). 
	\label{viscosity_condratio}
	\end{equation}
	$h_0(r_+)$ denotes the horizon value assumed by $h^y_x(r)$.
	A similar result obtains in isotropic backgrounds with momentum relaxation \cite{Hartnoll2016,Burikham2016,Alberte:2016xja}.
	$h_0(r_+)$ originates from the simultaneous breaking of rotations and translations along $y$ caused by the massless scalar. Since it has a non-trivial radial evolution, we expect that it will differ from unity as temperature decreases, i.e. as the system flows away from the UV AdS$_4$. {If the mass squared in Eq.\eqref{EOMdiss} is positive {(which is a sufficient condition for stability of the fluctuation spectrum, and always the case for the setups considered in this work)}, the metric fluctuation decreases towards the horizon-- see Ref.\cite{Hartnoll2016} for the analogue in the isotropic case. In anisotropic backgrounds where $m^2=0$, the viscosity-conductivity bound holds exactly also for the tensor element in  \eqref{viscosity_diss} -- see e.g. Ref.\cite{1604.01346}.} 
	
	\subsection{Analysis of the viscosity-conductivity bound}
	
	We would now like to discuss the temperature dependence of $h_0(r_+)$, both at high and low temperatures, and dependending on whether translations are broken explicitly or spontaneously. We first discuss the explicit case.
	
	\subsubsection{Explicit breaking of translations}

When $Y(\varphi(r\to0)) \to 1$ near the UV, the massless scalar induces a coordinate-dependent source, and translations are broken explicitly \cite{Andrade2014,Gouter_2014_mom_diss}.

	{At high temperature, i.e. $T\gg a$, the black hole horizon gets closer to the asymptotic region, and the geometry can be approximated by  $\text{AdS}_4$-Schwarzschild. The massless scalar $\psi$ sources deviations from isotropy by inducing $a^2$ corrections on the metric. Since the source in the Eq.\eqref{EOMdiss} depends on $a^2$, the second order correction to $h^y_x$ is determined by the isotropic background \cite{1904.11445}. {Eq.\eqref{viscosity_diss} becomes}
		\begin{equation}
			4\pi \frac{\sigma_{yy}}{\sigma_{xx}} \frac{\eta_{xyxy}}{s}= 1 - 2\int_0^{r_+} \frac{g_{rr}}{\sqrt{-g}}\left( \int_{r_1}^{r_+} \sqrt{ g_{rr} g_{tt}} \, Y(\varphi)\right)_{a=0} a^2 + O(a^4).
			\label{viscosity_high_T}
		\end{equation}
	 {The integral is positive-definite and the viscosity violates the lower conductivity-bound \eqref{eq:bound}. This effect has been related to the positivity of the graviton mass in Ref.\cite{Hartnoll2016}.}
	
	{ Using AdS-Schwarzschild background, we obtain: }
	\begin{equation}
	\label{boundESBSchwAdS}
		4\pi \frac{\sigma_{yy}}{\sigma_{xx}} \frac{\eta_{xyxy}}{s}  \approx 1 -  c_\text{esb} \left(\frac{a}{T}\right)^2,
	\end{equation}
	with $c_\text{esb}  = ({9 \log 3 -\sqrt{3} \pi})/{16 \pi^2} \approx 0.0281553$, which is similar to the results in the isotropic case, Ref.\cite{Hartnoll2016}.

	We can also discuss the temperature dependence of $h_0(r_+)$ at low temperatures. First, we discuss the case where the massless scalar $\psi$ vanishes faster than other bulk fields towards the extremal horizon. Then, the IR endpoint enjoys rotation and translation symmetries, which are broken only through an irrelevant deformation sourced by $\psi$, see also \cite{Davison:2018ofp,Davison2018}. These scaling solutions are discussed in appendix \ref{app:irraxion}. Since $\psi$ sources an irrelevant deformation, the formula \eqref{viscosity_high_T} still applies and we obtain
		\begin{equation}
	\label{boundESBIRz=1}
		4\pi \frac{\sigma_{yy}}{\sigma_{xx}} \frac{\eta_{xyxy}}{s}  \approx 1 -  c^{ir}_\text{esb} \left(\frac{a}{T^{\Delta_a}}\right)^2,
	\end{equation}
	where $\Delta_a<0$ is the infrared scaling dimension of $a$.
	
	Alternatively, the translation/rotation breaking field $\psi$ can source a marginal deformation at $T=0$. In this case, there is no notion of momentum, although of course we can still compute the response to shear strain using the Kubo formula. But then the object we are computing does not have the interpretation of a shear viscosity.
	Its temperature dependence follows from an asymptotic analysis near the boundary of the IR region and yields:
	\begin{equation}
	\frac{\sigma_{yy}}{\sigma_{xx}}\frac{\eta_{xyxy}}{s} \sim   T^{\frac{\delta_0-2(\phi-1)}{z} \left(-1+\sqrt{1+ 	 		
			\left(\frac{2 a L}{\delta_0-2(\phi-1)}\right)^2
		} \right) }. 
		\label{boundESBIRz!=1}
	\end{equation}
	The sign of the exponent is not fixed, hence the tensor element can vanish or diverge -- for details on the parameter range see Appendix \ref{single_axion_model}.
	This result is still valid when two massless scalars are taken into account  \eqref{scaling_solution}. The isotropic limit of this last case is consistent with Ref.\cite{Ling_2016,Ling_2017} at charge neutrality. 
\\

\subsubsection{Spontaneous breaking of translations}	
	When $Y(\varphi(r\to0)) \to \varphi^2$, the massless scalar and the dilaton can be rearranged as the phase and the norm of a single complex scalar field $\Phi$ -- see e.g. Ref.\cite{Donos2014,Amoretti:2017axe,Amoretti2018_CDW}. The model then describes a CFT deformed by a relevant complex coordinate-dependent operator, determined by the asymptotic expansion $\Phi(r\to 0,x)= \Phi_{s}(x)\, r^{3-\Delta_\varphi}+\Phi_{v}(x)\, r^{\Delta_\varphi}$. Picking appropriate UV boundary conditions for $\Phi$ allows translations to be broken spontaneously (SSB), restoring the hydrodynamic meaning of the viscosity.\footnote{Similar results would also obtain in so-called holographic massive gravity models, \cite{Alberte:2017oqx}.} In such a setup, $\eta$ can be directly extracted from the Kubo formula Eq.(\ref{viscosity_scaling_section}), see e.g. Ref.\cite{Delacretaz:2017zxd}.
	
	Since we are interested in $O(a^2)$ corrections at high temperatures, it is enough to consider the dilaton as a probe field in the $\text{AdS}_4$-Schwarzschild spacetime. Solving its decoupled equation of motion for $\Delta_\varphi=2$,\footnote{Using other values of the scaling dimension presents no conceptual obstacle, but the solution in these other cases cannot be obtained analytically.} the solution reads:
	\begin{equation}
		\Phi(r)=  r \left[ {_{2}F_1}\left(\frac{1}{3},\frac{1}{3};\frac{2}{3};\frac{r^3}{r_+^3}\right)  \Phi_s
		+ \,\,{_{2}F_1}\left(\frac{2}{3},\frac{2}{3};\frac{4}{3};\frac{r^3}{r_+^3}\right) {r} \, \Phi_v
		\right] 
	\label{full_dilaton_solution}
	\end{equation}
	where the $F$s are hypergeometric functions.
We wish to impose horizon regularity in this solution $\left[\Phi_s \Gamma^3 \left(\frac{2}{3}\right)+9 {\Phi_v} r_+ \Gamma^3 \left(\frac{4}{3}\right)\right]\log \big(1-\frac{r}{r_+}\big)=0$, 
 which yields a linear relation between the asymptotic coefficients 
	\begin{equation}
		{\Phi_v}=-\frac{12 \pi ^{3/2}}{{r_+} \Gamma^3\left(\frac{1}{6}\right)} \,{\Phi_s}.
		\label{linear_relation_coeff}
	\end{equation}
This implies that standard Dirichlet boundary conditions cannot consistently be imposed, and SSB cannot be realized in the standard way by setting $\Phi_{s}=0$. Indeed, it is well-known that when the squared-mass of the complex scalar lies in the window $[-9/4,-5/4]$, we can choose mixed boundary conditions for which the field is dual to an operator with $\langle\mathcal{O}\rangle = \Phi_s$ sourced by $J = - \Phi_v - F'(\Phi_s)$ -- see eg Refs.\cite{Witten:2001ua,Caldarelli2017}. $F$ is a polynomial whose degree lies in the interval $[2,3/(3-\Delta_\varphi)]$. In the following we choose the mass such that $\Delta_\varphi = 2$, and fix the form of $F$ by setting SSB conditions $J=0$. In particular, we find $F(\Phi_s) \propto \Phi_s^2$ as in Eq.\eqref{linear_relation_coeff}. As pointed out in Ref.\cite{Caldarelli2017}, the mixed boundary conditions generate an extra contact term in the dual stress-energy tensor. Since this contact term is real, it does not change the imaginary part of the retarded Green's function nor the shear Kubo formula for the shear viscosity Eq.\eqref{viscosity_scaling_section}. As such, previous relations such as Eq.\eqref{viscosity_high_T} continue to hold. 

We note that the leading deviations in \eqref{viscosity_high_T} appear at $O(\Phi_s^2 a^2)$. This means we do not need to consider the backreaction of the dilaton or the massless scalars on the metric, which would source higher order terms. Evaluating the integral on the isotropic background, we find: 
	\begin{equation}
	4\pi \frac{\sigma_{yy}}{\sigma_{xx}} \frac{\eta_{xyxy}}{s}  \approx 1 -  c_\text{ssb}\Big(\frac{\Phi_s}{T}\Big)^2  \left(\frac{a}{T}\right)^2,
	\end{equation}
with $c_\text{ssb} \approx 0.000435607$. We observe that compared to the explicit breaking case \eqref{boundESBSchwAdS}, violations of the bound are further suppressed by extra powers of $T$ and will generally become sizable for lower temperatures than in the explicit breaking case.\\
At low temperatures and under the same assumptions for the scalar couplings \eqref{coupling_IR} as in the explicit breaking case, the same temperature dependences are obtained, both for the irrelevant \eqref{boundESBIRz=1} and marginal cases \eqref{boundESBIRz!=1}. We emphasize that since translations are broken spontaneously, the shear viscosity remains a well-defined hydrodynamic coefficient that can be computed via the usual shear Kubo formula. 
	
	In any case, the viscosity-conductivity bound stated through the scaling analysis is holographically realized at least for one of the $\eta/s$-tensor elements.

	\section{Holographic analysis of the charge-diffusivity bound}
	The charge diffusivity in the $\alpha$-direction is determined by the electrical conductivity and the charge susceptibility via the Einstein relation
	{ $D_{\text{c},\alpha}=\sigma_{\alpha\alpha}/{\chi}_{\rho}$.} In section \ref{scaling_arguments} we demonstrated that the
	combination 
	\begin{equation}
	X_{\alpha}=\frac{k_{{\rm B}}TD_{c,\alpha}}{\hbar v_{\alpha}^{2}}
	\end{equation}
	has scaling dimension zero, which suggests that it approaches at low
	temperatures a universal value. In the subsequent sections we will
	use  Eq.(\ref{eq:conduct holo}) for the conductivity, obtained through the holographic approach and determine, within the same theory, the charge susceptibility
	and the butterfly velocity of the system.  Without loss of generality we set $\zeta=0$, as
	in the charge neutral case the exponent of $Z_\text{IR} = e^{\zeta \varphi}$ is not constrained -- see Appendix \ref{single_axion_model}.
	We then obtain the result that 
	\begin{equation}
	X_{\alpha}=\frac{1}{2\pi}\frac{1+\phi-\theta}{1+\phi-z} 
	\end{equation}
	independent	on the space direction $\alpha$ leads to Eq.(\ref{eq:Diffusivity aniso}).
	\subsection{Analysis of the diffusivity}
	An important ingredient for the bound on the diffusivity in Eq.(\ref{eq:diffusivity})
	is the isothermal charge susceptibility {$\chi_{\rho}=(\partial\rho/\partial\mu)_{T}$}. In order to derive the correspondent holographic relation, we formally solve Maxwell equations (\ref{maxwell_conserving}): 
	\begin{equation}
	A_t(r) = A_t (r_+) -\rho \int_{r_+}^r  \frac{dr}{\sg Z(\varphi) g^{rr}g^{tt}}\,.
	\end{equation}
	As mentioned, $A_t$ yields the chemical potential near the boundary and vanishes at the horizon, therefore
	{
	\begin{equation}
	{\chi}_{\rho}^{-1} =  \int_{0}^{r_+} \frac{dr}{\sg Z(\varphi) g^{rr}g^{tt}},
	\label{inverse_compressibility}
	\end{equation}
}	
	see also Ref.\cite{Iqbal2009}. {Due to the non locality of the above formula, the integral can only be worked out by explicitly solving the RG flow from the boundary to the horizon.
	Keeping in mind that $r_+\propto T^{-{1}/{z}}$, we observe that the near horizon geometry contribution scales as $T^{-{\Delta_\chi}/{z}}$. Within a low temperature analysis, this is the dominant term if ${\Delta_\chi}/{z}>0$ and the charge diffusion is uniquely controlled by the IR physics, in accord with the isotropic analysis of Ref.\cite{Blake2016,Blake2016b}. In this case we obtain
}
	{
	\begin{equation}
	{\chi}^{-1}_{\rho} = -\frac{L}{\Delta_\chi}\frac{r^{\Delta_\chi}}{Z(\varphi)}\bigg|_{r_+}.
	\end{equation} 
	}
	We can alternatively Taylor-expand the integrand $i(r)$ of (\ref{inverse_compressibility}) near the horizon. From the IR scaling behavior follows the recursion rule
	\begin{equation}
	i^{(n)} (r)= \frac{(-1)^n}{r^n} \left[\prod_{k=1}^{n} (k-\Delta_\chi)\right] i(r),
	\end{equation}
	$i^{(n)}(r)$ denotes the $n$-th radial derivative of $i(r)$. Plugging this expression into the Taylor expansion we find
	\begin{equation}
	i(r) = i(r_+) \sum_{n=0}^{\infty} {{n-\Delta_\chi}\choose{n}} \left({1-\frac{r}{r_+}}\right)^n
	\end{equation}
	Performing the binomial series we obtain $
	i(r) = i(r_+) ({{r}/{r_+}})^{\Delta_\chi-1}$, which yields the same result as the previous analysis.
	\smallskip\\
	The susceptibility together with the holographic conductivities (\ref{eq:conduct holo}) yields the diffusion constants
	\begin{equation}
	D_{\text{c},\alpha} 
	= -\frac{L}{\Delta_\chi} \frac{r^{\theta-z}}{g_{\alpha\alpha}(r)}\bigg|_{r_+}.
	\label{diffusivity}
	\end{equation}
	The above results are still valid in the $\zeta\neq0$ case.
	
	\subsection{Analysis of the butterfly velocity in anisotropic systems}
	Following Ref.\cite{Shenker2014}, we determine the butterfly velocity for an anisotropic holographic system using a shock-wave analysis. As mentioned in the introduction, the butterfly velocity can be thought of as the velocity of growth of out-of-time-order correlation functions of local operators.
	Holographically, it can be calculated from the back-reaction of the
	metric due to a massless particle falling towards the horizon of the black
	hole. The velocity of growth of this back-reaction can then be identified
	as the butterfly velocity. 
	
	For the subsequent analysis it is convenient to use Kruskal-coordinates
	\begin{equation}
	uv  = - e^{\gamma'(r_+) r_\star(r)}, \quad
	u/v = - e^{-\gamma'(r_+) t},
	\end{equation}
	where $\gamma'$ denotes the radial derivative. 
	$uv=0$ and $uv=-1$ correspond to the horizon and to the boundary respectively -- see Fig.\ref{Kruskal}. The anisotropic metric (\ref{eq:metric}) takes the form 
	\begin{equation}
	ds^{2}  =  -g_{uv}\left(uv\right)du dv\nonumber  + \sum_\alpha g_{\alpha\alpha}\left(uv\right)dx_\alpha^{2}.
	\end{equation}
	Next we perturb the system by adding $\delta T_{uu}\propto Ee^{2\pi Tt_{w}}\delta(u)\delta(x)\delta(y)$
	to the holographic stress-energy tensor, which represents a particle of energy $E$ released at the left boundary at time $t_{w}$ in the past and propagating
	towards the $u=0$ horizon \cite{Shenker2014,Sfetsos1995}. The perturbed
	metric can then be expressed in the following shock-wave form 
	\begin{eqnarray}
	ds^{2} & = & -g_{uv}\left(uv\right)du dv+g_{uv}\left(uv\right)h(x,y)du^{2}\nonumber \\
	& + & \sum_\alpha g_{\alpha\alpha}\left(uv\right)dx_\alpha^{2}.
	\end{eqnarray}
	The equation of motion for $h\left(x,y\right)$ follows from Einstein equations at near the $u=0$ horizon:
	\begin{equation}
	\left(
	\sum_\alpha 
	\frac{\partial_\alpha^2}{c_\alpha^{2}}
	-m_{h}^{2}\right)h(x,y)=b\delta(x)\delta(y),
	\label{shockwave_eq}
	\end{equation}
	with $c_\alpha = \sqrt{g_{\alpha\alpha}(0)}$, $ b\propto Ee^{2\pi Tt_{w}}/ {g_{uv}(0)}$ and mass
	\begin{equation}
	m_{h}^{2}=
	{\frac{1}{g_{uv}}}\frac{\partial \log(g_{xx} g_{yy})}{\partial (uv)}\Bigg|_{u=0}.
	\end{equation}
	\begin{figure}[t]\centering
		\includegraphics[scale=0.31]{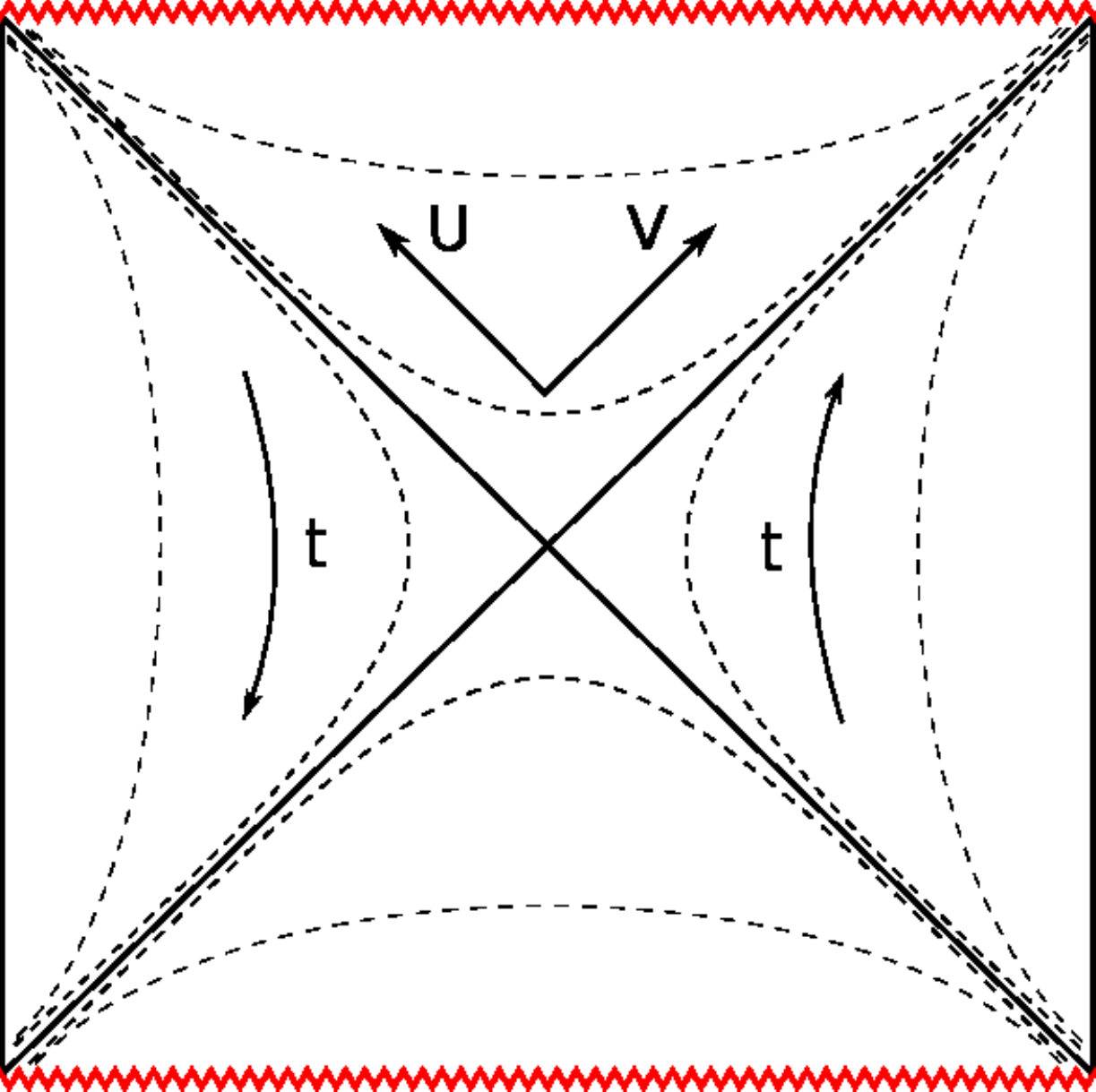}
		\caption{Black hole AdS space in Kruskal coordinates. The $(x,y)$-space,
			which is attached at each point, is not shown. The boundary (left/right
			edge) is located at $uv=-1$, the horizon (diagonal lines) is located
			at $uv=0$ and the singularity (upper/lower edge) is located at $uv=1$.
			Dashed lines represent surfaces of constant $u$. Time $t$ is flowing
			upwards in the right wedge and downwards in the left wedge.}
		\label{Kruskal}
	\end{figure}
	\\
	Eq.(\ref{shockwave_eq}) is consistent with the isotropic case of Ref.\cite{Blake2016b}. The solution can be expressed in terms of the 0th modified Bessel function of the second kind $\mathcal{K}_{0}$ as
	\begin{equation}
	h(x,y)\propto-\frac{b\, c_{x}c_{y}}{2\pi}\mathcal{K}_{0}
	(m_h\varrho),
	\end{equation}
	where  $\varrho^2={c_x^2 x^{2}+c_y^2 y^{2}}$. 
	At large values of $\varrho$, i.e. at large spatial distances, this gives 
	\begin{equation}
	h(x,y)\propto \frac{1}{{\sqrt{\varrho}}}
	\exp \left[{2\pi T\left({ t_{w}-\frac{m_{h}}{2\pi T} \varrho }\right)}\right].
	\end{equation}
	From the exponent we can extract the direction-averaged scale for the velocity
	\begin{equation}
	\bar{v}_{\mathrm{B}} = 
	\frac{2\pi T}{m_h}.
	\end{equation}
	In order to switch to the original system of coordinates, we use the identity $uv \,g_{uv}(uv)= g_{tt}(r)/\partial_r g_{tt}(r_+)^2$ and obtain
	\begin{equation}
	\bar{v}_{\mathrm{B}}^2
	= - {\frac{2\pi T L}{d_\text{eff}-\theta}}r_+^{\theta-z}.
	\end{equation}
	To determine the butterfly velocity along $x$,
	we consider the case where $y=0$ and we move in the $x$-direction.
	This gives 
	\begin{equation}
	\frac{\varrho}{\bar{v}_{\mathrm{B}}}=
	\frac{c_x|x|}{\bar{v}_{\mathrm{B}}}
	\equiv\frac{|x|}{v_{\mathrm{B},x}}.
	\end{equation}
	It then follows for the velocity along the $\alpha$-direction
	\begin{equation}
	v_{\mathrm{B},\alpha}=\frac{\bar{v}_{\mathrm{B}}}{\sqrt{g_{\alpha\alpha}(r_+)}},
	\end{equation}
	in accord with Ref.\cite{Jeong2018,Pedraza2018,1805.01470,1610.02669,1710.05765,Blake:2017qgd}.
	This result violates the upper bound of the isotropic case pointed out in Ref.\cite{1612.00082}, consistently with Ref.\cite{1708.07243,1811.06949}
	
	Considering the ratio between the diffusion constant (\ref{diffusivity}) and the square
	butterfly velocity we finally obtain
	\begin{equation}
	\frac{D_{\text{c},\alpha}}{v_{\text{B},\alpha}^{2}}=
	\frac{d_\text{eff}-\theta}{\Delta_\chi}\frac{1}{2\pi T},
	\end{equation}
	which yields the result Eq.(\ref{eq:Diffusivity aniso}) for the diffusivity bound.
	
	\section{Conclusions \label{sec:ccl}}
	{Motivated by previous results in anisotropic Dirac systems \cite{Link2018}, }in this paper we analyzed transport coefficients at a quantum Lifshitz point in the strong coupling limit, using scaling arguments and exploiting the duality between quantum field theories and gravity theories. We  have focused on particle-hole symmetric theories at charge neutrality which admit a gravitational dual description. We have shown that bounds on transport coefficients of the isotropic case can be generalized to the anisotropic one. 
	\\
	We analyzed the behavior of several observables after a spacetime dilatation, emphasizing that the scale dimensionless ones must approach a constant value for low temperatures. It turned out that some elements of the  $\eta/s$-tensor have a nonzero dimension while the diffusivity still exhibits the scaling of the rotational invariant case. In order to address the former, we included the electric transport, multiplying the ratio by a specific combination of conductivities such that the dimension of the resulting quantity is zero.
	\medskip\\
	Within the Einstein-Maxwell-dilaton model considered, translational symmetry is broken along the $y$ direction by a massless scalar in the bulk with a bulk profile linear in $y$. Thus, the $x$-component of the momentum
	is still conserved and $T_{\alpha x}$ continues to be the current
	of a conserved quantity. Therefore, the viscosity tensor elements $\eta_{\alpha x\beta x}$
	maintain their meaning as hydrodynamic transport coefficients. Since
	we can find solutions of the field equations that yield either $\phi<1$
	or $\phi>1$, we can always construct an anisotropic geometry that
	violates the isotropic viscosity bound for at least one tensor element, while fulfilling the generalized bound given in Eq.\eqref{eq:at the bound anisotr.}.}

In the direction where translations are broken, momentum relaxes at a rate $1/\tau_\text{mr}$. {In a holographic system with slow momentum relaxation} $1/\tau_\text{mr}\ll\Lambda$, where $\Lambda$ is a UV cutoff, there is a range of intermediate times $1/\tau_\text{mr}\lesssim t\ll\Lambda$ where momentum is approximately conserved. In this regime, the viscosity can be defined from the shear Kubo formula, yet is still found to violate the viscosity-to-entropy-density-ratio bound. 
Alternatively, it is likely that the diffusivity of transverse momentum is a better quantity to bound: the results of Ref.\cite{Ciobanu:2017fef} show that it obeys a bound of the kind \eqref{eq:diffusivity}, with the speed of light as the characteristic velocity.

{We also discussed the effects of translational symmetry breaking on the conductivity-viscosity bound, both in the explicit and spontaneous setups. In this latter case, deviations from the bound are more suppressed for high enough temperatures than in the explicit one.}

Differently from the other quantities, the diffusivity is not solely given by data on the horizon and is expressed through an integral over the radial direction. Although we do not have the full expression of bulk fields, we have derived a near horizon formula for the compressibility, and could relate the diffusion constant to the horizon data in a simple fashion. Indeed, the near IR geometry dominates at low temperature \cite{Blake2016,Davison:2018ofp,Davison2018}.
On the other hand we have calculated the butterfly velocities by moving to the Kruskal system of coordinates and using a generalization of the shock-wave technique. We have computed the proportionality factor between the diffusivity to the square butterfly velocity ratio and the inverse temperature, finding that it can be expressed in terms of the critical exponents $z,\phi$, and $\theta$.
\\
{It is well-known that the KSS bound can be violated by terms containing more than two derivatives of the metric (see Ref.\cite{Cremonini:2011iq} for a review), which capture finite 't Hooft coupling corrections. A possible extension of our work would be to consider the effects of higher derivative terms involving the massless scalars, along the lines of Ref.\cite{Baggioli2017}}. Moreover, particle-hole symmetry breaking could be taken into account as well.
\\
In our work, the anisotropy was sourced by a massless scalar that also breaks translations. As we have discussed, only the bound involving the transverse viscosity is preserved. In Ref.\cite{1604.01346}, isotropy is broken by one of the spatial components of the gauge field acquiring a vev. In this setup, translations are unbroken and viscosities are still well-defined hydrodynamic transport coefficients. Their results, specifically equation (14), show that the bound \eqref{eq:bound} holds in this holographic setup. It would be interesting to further investigate such systems, as well as different sources of spontaneous anisotropy, \cite{Donos:2012gg,Hoyos:2020zeg}.\\
Thus, we conclude that the transport properties of a strongly-interacting many-body system near a quantum Lifshitz point can be efficiently described using holographic methods and requires a generalization of the viscosity bound obtained in isotropic theories.

\begin{acknowledgments}
	We are grateful to A. Amoretti, S. A. Hartnoll, E. I. Kiselev, N. Maggiore, N. Magnoli, B. N. Narozhny, and K. Schalm
	for stimulating discussions. We thank European Commission’s Horizon 2020 RISE program Hydrotronics (Grant Agreement 873028) for support.
	BG has been supported during this work by the European Research Council (ERC) under the European Union's Horizon 2020 research and innovation programme (grant agreement No 758759).
\end{acknowledgments}

\appendix
\section{ Scaling of the viscosity tensor}
\label{Alternative_viscosity}
In this appendix we offer an alternative derivation of the scaling
dimension, Eq.(\ref{eq: viscosity dimension}) of the viscosity tensor.
The analysis leads to results identical to those presented in Section
\ref{scaling_arguments} of the paper. 
\\
Since the viscosity tensor describes the linear response to the temporal
change of an externally-applied strain field, we can also define it using the strain generators $\mathcal{J}_{\alpha\beta}.$
The strain generators describe the deformation of the coordinate systems
due to an applied external strain and are given by \cite{Bradlyn2012,Link2018b} $\mathcal{J}_{\alpha\beta}=x_{\alpha}k_{\beta}+\frac{i}{2}\delta_{\alpha\beta}$.
Hence, the viscosity tensor is defined as
\begin{equation}
\eta_{\alpha\beta\gamma\delta}(\omega)=\omega\,\mathrm{Im}\chi_{\alpha\beta\gamma\delta}^{(\mathcal{J})}(\omega)\,,
\end{equation}
with  $\chi_{\alpha\beta\gamma\delta}^{(\mathcal{J})}(\omega)$ being the  Fourier transform of $\chi^{\mathcal{J}}_{\alpha \beta \gamma \delta}(t) = - i \theta(t) \langle [ J_{\alpha \beta}(t),J_{\gamma \delta}(0)] \rangle$,  where $J_{\alpha \beta}$ is the density of the strain generator $\mathcal{J}_{\alpha \beta}$. In order to obtain the scaling dimension of the correlation function, we assume for the strain generator density the same dimensionality as the particle density $\Delta_{\rho}=d_{eff}$ times the scaling dimension of the momentum coordinates $k_\beta$ , $k_\delta$ and the spatial coordinates $x_\alpha$, $x_\gamma$, which have the dimensionality of the inverse momentum. We find for the correlation function of the two strain generators
\begin{equation}
\chi_{\alpha\beta\gamma\delta}^{(\mathcal{J})}(k_{\perp},\boldsymbol{k}_{\parallel},\omega)=b^{-\Delta_{\mathcal{J},\alpha\beta\gamma\delta}}\chi_{\alpha\beta\gamma\delta}^{(\mathcal{J})}(b^{\phi}k_{\perp},b\boldsymbol{k}_{\parallel},b^{z}\omega)
\end{equation}
with $\Delta_{\mathcal{J},\alpha\beta\gamma\delta}=d_{\mathrm{eff}}-z-\varphi_{\alpha}+\varphi_{\beta}-\varphi_{\gamma}+\varphi_{\delta}.$
Here we used the same notation as in the main paper, where $\varphi_{\alpha}=1$
if the $\alpha$-component is alinged along the direction of $\boldsymbol{k}_{\parallel}$
and $\varphi_{\alpha}=\phi$ for the direction of $k_{\perp}$. Using
$\Delta_{\mathcal{J},\alpha\beta\gamma\delta}$ allows us to determine
the scaling behavior of the viscosity tensor
\begin{equation}
\eta_{\alpha\beta\gamma\delta}(T,\omega)=b^{-\Delta_{\eta,\alpha\beta\gamma\delta}}\eta_{\alpha\beta\gamma\delta}(b^{z}T,b^{z}\omega)
\end{equation}
with
\begin{eqnarray}
\Delta_{\eta,\alpha\beta\gamma\delta} & = & \Delta_{\mathcal{J},\alpha\beta\gamma\delta}+z\notag\\
& = & d_{\mathrm{eff}}-\varphi_{\alpha}+\varphi_{\beta}-\varphi_{\gamma}+\varphi_{\delta}\:,
\end{eqnarray}
which is in agreement with Eq.(\ref{eq: viscosity dimension}) of
the main part of the paper.

\section{{IR models}}
\label{single_axion_model}
In order to analyze the IR metric (\ref{IR_scaling_model}), we derive the hyperscaling-violating solutions in the presence of both one and two massless scalar fields. It is worth to emphasize that the radial coordinate parameterizing the IR geometry \eqref{IR_scaling_model} does not coincide with the one in the UV region \eqref{AdS_metric}. \cite{Davison2018} 
To be specific, we consider the matter Lagrangian
\begin{equation}
{\cal L}_{\text{M}}= 
-
\frac{1}{2}\left(\nabla\varphi\right)^{2}+V_0 r^{2\kappa \delta}
-\sum_{\alpha=1}^p \frac{r^{2\kappa \lambda_\alpha}}{2}\left(\nabla\psi_\alpha\right)^{2}
-\frac{r^{2\kappa \zeta}}{4}F^{2},
\end{equation}
where $p$ is the number of massless scalars and $\psi_\alpha=a_\alpha x_\alpha$, with no index summation. In the $p=1$ case it reduces to (\ref{matter_lagrangian}).
\\
The effective dilaton potential (\ref{Veff}) looks like
\begin{equation}
\frac{V_\text{eff}(r)}{\sg} =
\frac{1}{2} \sum_{\alpha=1}^p a_\alpha^2 r^{2\Lambda_\alpha-\theta}-V_0 r^{2 \delta  \kappa},
\end{equation}
where $\Lambda_1=\kappa\lambda_x+\phi$ and $\Lambda_2=\kappa\lambda_y+1$.	
\subsection{Marginally relevant case}
In order to avoid radial dependences coming from the $a_\alpha$-terms, we set $2\Lambda _\alpha=2\kappa\delta +\theta$. 
This corresponds to take the massless scalars as marginal deformations of the IR fixed point. Furthermore, setting $\theta+2\delta\kappa =0$ yields a set of algebraic equations in both the cases $p=1,2$.
\\
Let us start with the one single massless scalar case $p=1$ -- we omit the subscript $\alpha=1$ everywhere. 
The solution to the field equations is given by:
\begin{eqnarray}
&&z=\phi, \quad 2\kappa\delta=-\theta, \quad \kappa\lambda =-1, \nonumber\\
&&4\kappa^2 =\theta^2  -2 \theta \phi + 2\phi -2 ,
\nonumber\\
&&L^2={(\theta -2\phi-1) (\theta -2\phi )}/{V_0} ,
\nonumber\\ 
&&a^2=\frac{2V_0 (1 -\phi)}{\theta -2\phi}.
\label{solution_single_axion_marginal}
\end{eqnarray}
Note how a low momentum dissipation limit ($a\to0$) always restores the isotropy of the system ($\phi=1$). 
In order to get a realistic solution, we demand the positivity of the squared quantities and the specific heat $c=T\,\partial_T s$. In addition, we require the vanishing of the line element in the IR at $T=0$, obtaining the following set of conditions:
\begin{eqnarray}
&&\theta <2 ,\,\,\, \theta ^2+2 \phi >2 \theta  \phi +2,\,\,\, \phi >1,
\\
&&\theta >2,\,\,\, \theta ^2+2 \phi >2 \theta  \phi +2,\,\,\, \theta  \phi <\phi ^2+\phi.
\label{consitency_conditions_1axion_marginal}
\end{eqnarray}	
In the former the IR is at $r=\infty$, in the latter at $r=0$. 
The null energy condition (NEC) turns out to be fulfilled. 

In the $p=2$ case we find
\begin{eqnarray}
&&2 \kappa \delta =-\theta, \quad  \kappa\lambda _x=-\phi , \quad \kappa \lambda _y=-1,
\nonumber\\
&&4 \kappa ^2=\theta(\theta-2z)-2\phi(\phi-z)-2(1-z),
\nonumber\\
&&L^2={(\theta-2 z ) (\theta-\phi-z -1 )}/{V_0} , 
\nonumber\\
&&a_x^2=\frac{2 V_0 (\phi-z )}{\theta-2 z },
\quad
a_y^2=\frac{2 V_0 (1-z)}{\theta-2 z },
\label{scaling_solution}
\end{eqnarray}
which reproduces Eq.\eqref{solution_single_axion_marginal} in the $a_x=0$ case. The consistency conditions follows from analogue considerations and are depicted in Fig.\ref{Parameters3d} -- the NEC is automatically satisfied. 
\begin{figure}[t]\centering
	\includegraphics[width=.5\textwidth]{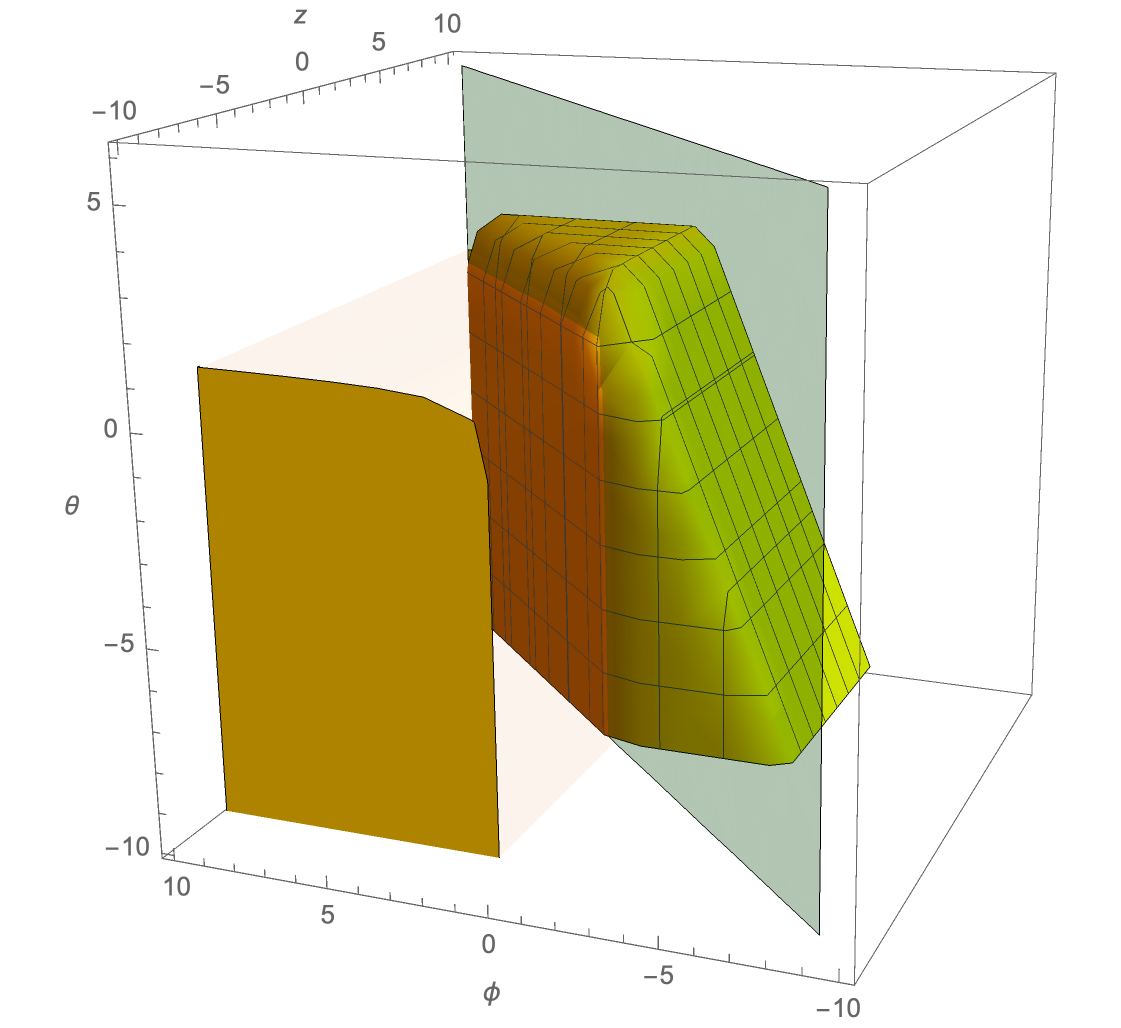} 	
	\includegraphics[width=.42\textwidth]{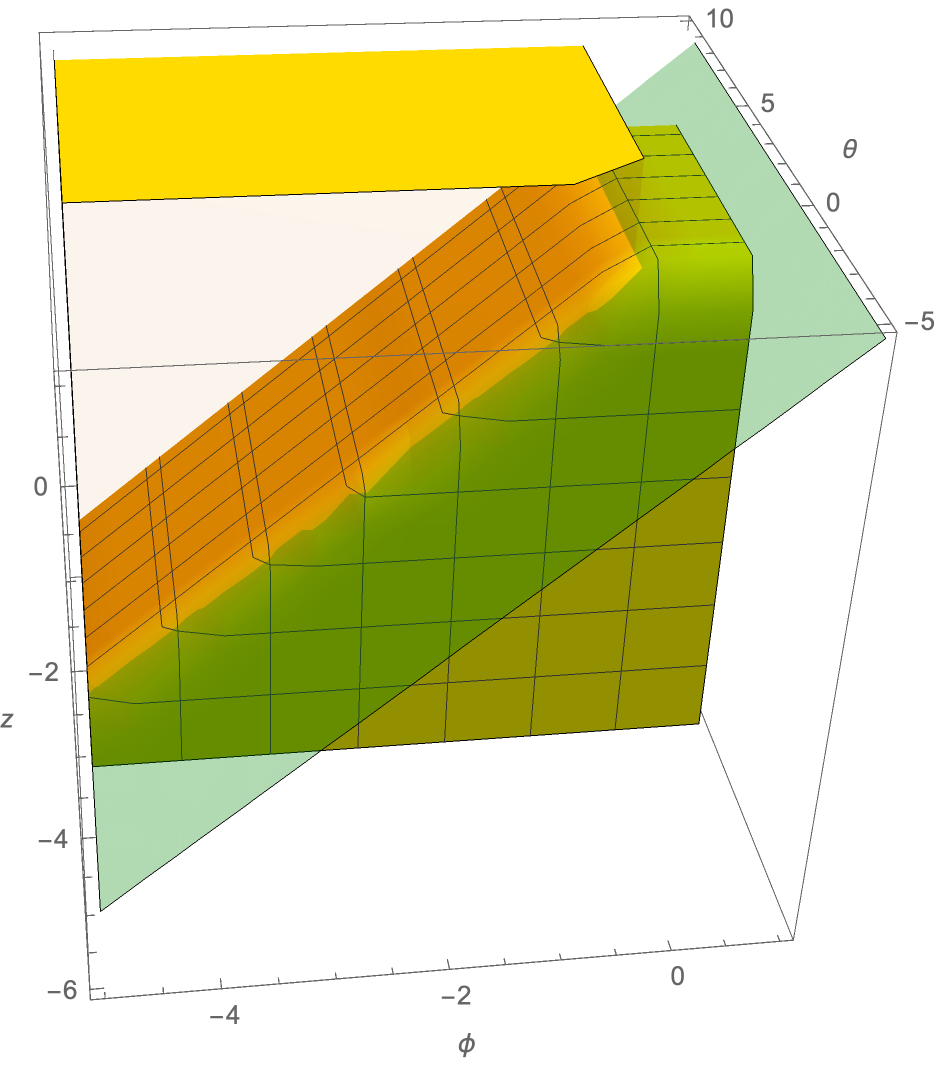} 
	\caption{Parameter space for the double marginal massless scalars solution. The intersection with the plane $z=\phi$ coincides with  \eqref{consitency_conditions_1axion_marginal}.}
	\label{Parameters3d}
\end{figure}
Even in this case, sending the momentum dissipation to zero restores the isotropy of the system.
\\
One can easily check that the above solution reproduces the single massless scalar one when $a_x=0$.

\subsection{Irrelevant case\label{app:irraxion}}
Now we wish to investigate the $p=1$ case, where the massless scalar acts as an irrelevant deformation of the IR endpoint. Details on the $p=2$ mixed case can be found in Ref.\cite{Davison2018,Jeong2018}. We firstly determine the solution when $a=0$ and then consider perturbations of the form:
\begin{equation}
\Phi = \Phi_{a=0} \left( 1+ c_\Phi a^2 r^{2\Delta_a}\right).
\end{equation}
$\Phi$ stands for the metric elements or the dilaton field, and $c_\Phi$ are numerical coefficients that follow from the $\mathcal{O}(a^2)$ fields equations. Such corrections are expressed in terms of $a^2$ as the massless scalar enters quadratically the field equations. 
The leading solution is given by
\begin{eqnarray}
&&z=\phi=1, \nonumber\\
&&4\kappa^2 =\theta  (\theta -2),
\nonumber\\
&&L^2={(\theta -2) (\theta -3)}/{V_0},
\label{solution_single_axion_irrelevant}
\end{eqnarray}
provided that $\theta+2\delta\kappa =0$. Moreover we obtain 
\begin{equation}
\Delta_a = 1+\kappa\lambda
\end{equation}
in accord with Ref.\cite{Davison2018}.\footnote{Notice the different normalization $2\kappa_{here}=\kappa_{there}$} The consistency conditions read 
\begin{eqnarray}
\theta<0,\quad \Delta_a < 0,
\label{consitency_conditions_1axion_irrelevant}
\end{eqnarray}
given that the IR is located at $r\to+\infty$.

\section{The holographic dual of out-of-time-order correlation functions}

In the context of the butterfly velocity, we consider out-of-time-order correlation functions (OTOCs) of the form $C(\vec{x},t_w) = -\langle [A(\vec{x},t_w),B(0,0)]^2 \rangle$,  where $A$ and $B$ are hermitian local operators. In order to translate such functions to the holographic language, it is convenient to regularize them by rotating one of the commutators halfway around the thermal circle  \cite{Maldacena2016}. This results in
\begin{equation}
C(\vec{x},t_w) = -\mathrm{tr}\left[ \bar{y} [A(\vec{x},t_w),B(0,0)] \bar{y} [A(\vec{x},t_w),B(0,0) \right],
\end{equation}
where $\bar{y}$ is the squareroot of the density matrix. Next, we introduce the thermofield-double (TFD) state
\begin{equation}
\ket{\beta} =\frac{1}{\mathcal{Z}^{1/2}}\sum_n e^{-\beta E_n/2}\ket{n}_\mathrm{L}\! \ket{n}_\mathrm{R},
\end{equation}
with the partition function $\mathcal{Z}$ and the inverse temperature $\beta$. This state lies in the product space of two copies of the Hilbert space and $\ket{n}_\mathrm{L}$ and $\ket{n}_\mathrm{R}$ denote energy Eigenstates with Eigenvalues $E_n$ in the respective copies. Operators acting on the two copies are defined as $ O_\mathrm{L} = O^\top \otimes {\bold 1}$ and $ O_\mathrm{R} = {\bold 1} \otimes O$. With these definitions, the regularized OTOC can be written as an expectation value in the TFD state, i.e.
\begin{align}
C(\vec{x},t_w) = -\bra{\beta}  [B_\mathrm{L}(0,0),A_\mathrm{L}(\vec{x},t_w)] 
\cdot[A_\mathrm{R}(\vec{x},t_w),B_\mathrm{R}(0,0)]\!\ket{\beta}.
\end{align}
Furthermore, we note that the TFD state is invariant under time translations generated by $H_\mathrm{tot}=H_\mathrm{R}-H_\mathrm{L}$.
\par To proceed, we need to investigate the transition amplitudes prepared by $\ket{\beta}$ in order to identify the spacetime connecting the L and the R system. For two given states $\ket{\xi}$ and $\ket{\zeta}$, these transition amplitudes are given by 
\begin{equation}
\bra{\xi}_\mathrm{R}\! \braket{\zeta|_\mathrm{L}|\beta} \propto  \braket{\zeta|e^{-\beta H/2}|\tilde{\xi}},
\end{equation}
where the conjugate state $\ket{\tilde{\xi}}$ is defined such that $\braket{n|\tilde{\xi}} = \braket{\xi|n}$ for all states $\ket{n}$. This definition is only well-defined if the states $\ket{n}$ are redefined by $\ket{n} \to e^{-i\mathrm{arg}\braket{\xi|n}} \ket{n}$ in order to make the scalar products real. Using the fact that the Hamilton operator $H$ is obtained from the Hamilton density by integrating over the position space $\mathcal{P}$, the transition amplitude shows that the L and R systems are connected by the spacetime
\begin{equation}
\mathcal{B} = [0,\beta/2] \times \mathcal{P}.
\end{equation}
According to the holographic dictionary, this spacetime is the boundary of its holographic dual. It was shown in  \cite{Maldacena2001}, that the holographic dual of the TFD state is given by a two-sided black hole spacetime. For simplicity, we will demonstrate this for the case of a one-dimensional position space $\mathcal{P}$, but the results hold in any dimension. We first consider a Euclidean black hole in three dimensions, whose metric can be written in the two equivalent forms
\begin{align}
d s^2 &= (r^2-r_+^2) d \tau^2 + \frac{1}{r^2-r_+^2} d r^2 + g_{xx}(r) d x^2, \\
d s^2 &= \frac{4}{(1-zz^\star)^2} d z d z^\star + g_{xx}(zz^\star) d x^2.
\end{align}
The coordinate $x$ is restricted to the position space $\mathcal{P}$ and the two expressions are related by $z=e^{\frac{2\pi}{\beta}(r_\star(r)-i \tau)}$ with the tortoise coordinate.  Here, $g_{tt}(r)=g_{\tau\tau}(r)=r^2-r_+^2$ and $g_{rr}(r)=1/(r^2-r_+^2)$. 

\par If $\tau$ is restricted to the Euclidean time interval $[0,\beta/2]$, the boundary of the Euclidean black hole is equal to $\mathcal{B}$. This can be achieved by cutting the spacetime along the $\mathrm{Im}(z)=0$ surface. Furthermore, the metric is invariant under time translations of the form $z \to z\cdot e^{-i\frac{2\pi}{\beta}\Delta\tau}$. Such time translations change the position of the $\mathrm{Im}(z)=0$ surface, but leave the distance between its two boundary points invariant.

\par What remains to be implemented, is the Lorentzian time invariance. We achieve this analytic continuation by introducing Kruskal coordinates $z=-v$ and $z^\star=u$, yielding
\begin{equation}
d s^2 = \frac{-4}{(1+uv)^2} d u d v + g_{xx}(uv) d x^2.
\end{equation}
This metric is invariant under Lorentzian time translations $u \to u\cdot e^{-\frac{2\pi}{\beta}\Delta t}$, $v \to v\cdot e^{\frac{2\pi}{\beta}\Delta t}$.
In order to continuously connect the Kruskal coordinate frame to the Euclidean black hole, we rewrite $u=t+w$ and $v=t-w$, giving
\begin{align}
d s^2 = \,\frac{-4}{(1+t^2-w^2)^2} d t^2 + \frac{4}{(1+t^2-w^2)^2} d w^2 
+ g_{xx}(uv) d x^2.
\end{align}
At $t=0$, this metric is equal to the Euclidean black hole at the $\mathrm{Im}(z)=0$ surface. Thus, we can glue the Kruskal extension of the Lorentzian black hole to the Euclidean black hole along these surfaces (see Fig.\ref{fig:GluedMetric}). The resulting spacetime is the holographic dual of the TFD state.

\begin{figure}
	\centering
	\includegraphics[scale=0.9]{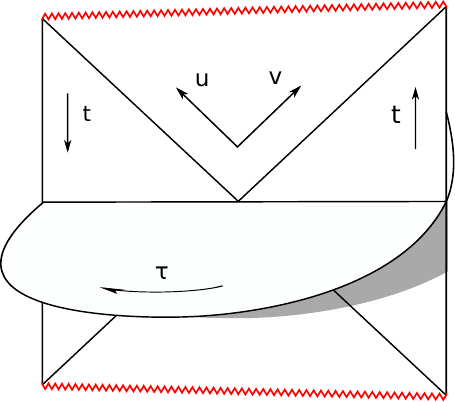}
	\caption{The plane represents the two-sided black hole in Lorentzian time. This is glued to the $\mathrm{Im}(z)=0$ surface of the Euclidean spacetime along the $t=0$ surface. The boundary regions L and R are separated along the thermal circle by a Euclidean time interval of length $\beta/2$. Lorentzian time flows forwards on the right boundary and backwards on the left boundary. The position space, which is attached at each point, is omitted.}
	\label{fig:GluedMetric}
\end{figure}

\par Now that we have demonstrated the duality between the TFD state and a two-sided black hole for a one-dimensional $\mathcal{P}$, we can return to a two-dimensional $\mathcal{P}$ and implement the effect of the OTOC. OTOCs are used as a measure for the butterfly effect, which describes how a microscopic effect (on UV energy scales) can become a macroscopic effect (on IR energy scales) at later times. This motivated Shenker and Stanford to propose that the OTOC can be modelled hoographically by a massless particle, which is thrown into the system at the boundary (the UV region) and has an effect at the horizon (the IR region) at a later time  \cite{Shenker2014}.

\par In order to quantify the effect of such a particle, we start with the action of a point particle, which can be written in the form
\begin{align}
\mathcal{S}[z^\rho] = \frac{1}{2} \int d \lambda\, e(\lambda) \left(\frac{1}{e(\lambda)^2}\frac{d z^\alpha}{d\lambda}\frac{d z^\beta}{d\lambda} g_{\alpha\beta}(z^\rho(\lambda)) - m^2\right).
\end{align}
Here, $z^\rho(\lambda)$ is a geodesic with affine parameter $\lambda$ and $e(\lambda)$ is a non-dynamical auxiliary field called 'Einbein'. In the massive case, this field is entirely fixed by the field equations and in the massless case, it can be interpreted as a gauge degree of freedom. The stress-energy tensor for a massless particle is given by
\begin{eqnarray}
T_{\mu\nu}(x^\rho) = \frac{2}{\sqrt{-g(x^\rho)}}\frac{\delta \mathcal{S}}{\delta g^{\mu\nu}} 
= \int d \lambda \frac{1}{e(\lambda)}\frac{d z_\mu}{d\lambda}\frac{d z_\nu}{d\lambda} \frac{\delta^{(4)}(x^\rho-z^\rho(\lambda))}{\sqrt{-g(x^\rho)}}.\qquad
\end{eqnarray}
A light-like infalling geodesic requires $dt/dr = - 1/\gamma(r)$. If the particle is inserted at the boundary at time $-t_w$ , the resulting geodesic is given by
\begin{equation}
(z^\mu(\lambda))^\top = (- r_\star(\bar{r}(\lambda))-t_w, \bar{r}(\lambda), 0, 0),
\end{equation}
where $\bar{r}(\lambda)$ denotes the radial coordinate at position $\lambda$. If $\lambda$ is identified with the radial coordinate, $e(\lambda)$ has units of inverse mass and can be identified as the inverse of the particle energy $E$. Such a parameterization may not in general be a solution of the geodesic equation. However, for a different parameterization, the resulting stress-energy tensor would only change by a global factor. It is thus sufficient to assume the $\bar{r}(\lambda)=\lambda$ case.

\par The stress-energy tensor is now given by
\begin{align}
T^{\mu\nu}(x^\rho) = E\frac{\delta(x)\delta(y)\delta(t + t_w + r_\star(r))}{\sqrt{-g(x^\rho)}}
\cdot\left(\frac{\delta^{\mu t}\delta^{\nu t}}{\gamma^2(r)} - \frac{\delta^{\mu t}\delta^{\nu r} + \delta^{\mu r}\delta^{\nu t}}{\gamma(r)} + \delta^{\mu r}\delta^{\nu r}\right).
\end{align}
At large $t_w$, after switching to Kruskal coordinates, the only non-vanishing component of the stress-energy tensor for a particle inserted at the right boundary is given by
\begin{equation}
T_{vv} \propto E e^{\frac{2\pi}{\beta}t_w} \delta(x)\delta(y)\delta(v).
\end{equation}
For a particle inserted at the left boundary, time is reversed ($t\leftrightarrow -t$), which is equivalent to $u\leftrightarrow v$. In this case, the stress-energy tensor is given by
\begin{equation}
T_{uu} \propto E e^{\frac{2\pi}{\beta}t_w} \delta(x)\delta(y)\delta(u).
\end{equation}
As discussed above, perturbing a two-sided black hole with this stress-energy tensor results in a shock wave, which can be identified as the OTOC.

\bibliographystyle{JHEP}
\bibliography{refs}

\end{document}